\documentclass[%
prx,
superscriptaddress,
 amsmath,amssymb,
 aps,
 reprint,
floatfix,
nofootinbib
]{revtex4-2}
\usepackage[dvipsnames]{xcolor}

\usepackage[normalem]{ulem}
\usepackage{wasysym}
\usepackage{verbatim}
\usepackage{mathtools,upgreek}
\usepackage{mathrsfs}

\usepackage{graphicx}
\usepackage{dcolumn}
\usepackage{bm}
\usepackage{hyperref}
\hypersetup{colorlinks = true, linkcolor = [rgb]{0.19411,0.51882,0.667058}, urlcolor = [rgb]{0.125490,0.29542,0.1647058}, citecolor = [rgb]{0.75882,0.37411,0.14117}}

\usepackage{setspace}
\usepackage{eurosym}
\usepackage{amssymb}

\usepackage{braket}

\usepackage{siunitx}

\graphicspath{{../images/}}
\usepackage[title]{appendix}
\usepackage{etoolbox,bm}

\patchcmd{\appendices}{\quad}{.\quad }{}{}

\newcommand{\nn}{\nonumber \\}

\usepackage{braket}
\usepackage[bottom]{footmisc}
\def\>{\rangle}
\def\<{\langle}

\DeclareMathOperator{\tr}{Tr}

\usepackage[bottom]{footmisc}
\usepackage{cleveref}
\begin{document}

\title{A quantitative comparison of amplitude versus intensity interferometry for astronomy}

\author{Manuel Bojer}
\email{manuelbojer6@gmail.com}
\affiliation{Institut f\"ur Optik, Information und Photonik, Universität Erlangen-N\"urnberg, 91058 Erlangen, Germany}

\author{Zixin Huang}
\email{zixin.huang@sheffield.ac.uk}
\affiliation{ Department of Physics \& Astronomy, University of Sheffield, UK }
\affiliation{Center for Engineered Quantum Systems, Department of Physics and Astronomy, Macquarie University, NSW 2109, Australia}

\author{Sebastian Karl}
\email{seb.karl@fau.de}
\affiliation{Institut f\"ur Optik, Information und Photonik, Universität Erlangen-N\"urnberg, 91058 Erlangen, Germany}

\author{Stefan Richter}
\email{stefan.michael.richter@fau.de}
\affiliation{Institut f\"ur Optik, Information und Photonik, Universität Erlangen-N\"urnberg, 91058 Erlangen, Germany}
\affiliation{Erlangen Graduate School in Advanced Optical Technologies (SAOT)}

\author{Pieter Kok}
\email{p.kok@sheffield.ac.uk}
\affiliation{ Department of Physics \& Astronomy, University of Sheffield, UK }

\author{J.\ von Zanthier}\email{joachim.vonzanthier@fau.de}
\affiliation{Institut f\"ur Optik, Information und Photonik, Universität Erlangen-N\"urnberg, 91058 Erlangen, Germany}
\affiliation{Erlangen Graduate School in Advanced Optical Technologies (SAOT)}

\begin{abstract}\noindent
Astronomical imaging can be broadly classified into two types. The first type is amplitude interferometry, which includes conventional optical telescopes and Very Large Baseline Interferometry (VLBI). The second type is intensity interferometry, which relies on Hanbury Brown and Twiss-type measurements. 
At optical frequencies, where direct phase measurements are impossible, amplitude interferometry has an effective numerical aperture that is limited by the distance from which photons can coherently interfere.
Intensity interferometry, on the other hand, correlates only photon fluxes and can thus support much larger numerical apertures, but suffers from a reduced signal due to the low average photon number per mode in thermal light.
It has hitherto not been clear which method is superior under realistic conditions.
Here, we give a comparative analysis of the performance of amplitude and intensity interferometry, and we relate this to the fundamental resolution limit that can be achieved in any physical measurement. 
Using the benchmark problem of determining the separation between two distant thermal point sources, e.g., two adjacent stars, we give a short tutorial on optimal  estimation theory and apply it to stellar interferometry. 
We find that for very small angular separations the large baseline achievable in intensity interferometry can more than compensate for the reduced signal strength. 
We also explore options for  practical implementations of Very Large Baseline Intensity Interferometry (VLBII).
\end{abstract}
\date{\today}
 
\maketitle

\section{Introduction}\noindent
Imaging takes a primary place in the instrumentarium of science. It has been responsible for some of the most dramatic discoveries, from Galileo's observations of the orbits of Jupiter's moons that were essential in overturning the Ptolemaic world view, to the imaging of the black hole in M87 by the Event Horizon Telescope (EHT) Collaboration that provided direct evidence for the existence of black holes and a stunning confirmation of Einstein's general relativity \cite{Akiyama1}. In microscopy, imaging has revolutionised our understanding of the natural world many times over, from Antonie van Leeuwenhoek's discovery of micro-organisms to superresolution imaging of large molecules.

It is well-known that the wave nature of light places a limit on the resolution that can be achieved in an image. The maximum resolution for a direct imaging system is given by Abbe's limit, which is determined by the wavelength of the light divided by the numerical aperture of the imaging apparatus \cite{lipson2010optical}. Consequently, much of the efforts in producing better images in stellar astronomy has been in creating telescopes with larger numerical apertures. The Extremely Large Telescope (ELT) in Chile is currently the largest optical and infrared telescope with a compound mirror of 39.3~m in diameter \cite{ELT_Construction_2014}. Moreover, the GRAVITY Collaboration at the Very Large Telescope (VLT) aims to collect and interfere optical signals from a number of telescopes, achieving an effective numerical aperture on the order of \SI{100}{\meter} \cite{2017_GRAVITY}. The Center for High Angular Resolution Astronomy (CHARA) collects coherently infrared light from six telescopes of \SI{1}{\meter} diameter, e.g., in the MIRC-X instrument, spanning an effective numerical aperture of \SI{300}{\meter} \cite{2020_Anugu, tenBrummelaar2005}.
Even more impressive, the EHT Collaboration used an array of existing telescopes all over the globe to create an effective numerical aperture the size of the Earth, with a resolution better than 20~$ \upmu$as. However, the latter method is restricted to lower frequency signals (the observed wavelength was 1.3~mm), since the telescopes have to be phase synchronized, i.e., each telescope must make a phase measurement of the electromagnetic field, in order to combine the many signals into a single coherent image \cite{Akiyama1}. At higher frequencies, detectors are not phase stable and fast enough, and we must find alternative ways to improve the resolution.

Recently, Nair and Tsang discovered that Abbe's diffraction limit is an artefact of the imaging system \cite{Nair}, and that in principle one can construct an imaging apparatus (with a finite numerical aperture) that can resolve two incoherent point sources at distances much lower than the wavelength of the light. 
Crucially, this does not require manipulation of the object, as is common in many modern super-resolution techniques \cite{schermelleh2019super}. The reason is that the quantum state of light  contains the information about the separation of the sources, but in the detection plane of the imaging system this information is enclosed in the phase instead of the intensity. Changing the measurement configuration can unlock this information in principle. It was subsequently shown by Lupo \emph{et al.}\ that imaging of $N$ incoherent point sources can be achieved equally optimally as in \cite{Nair} using again  only linear optics  and phase sensitive interferometry \cite{PhysRevLett.124.080503}.

Working towards a less involved and more general-purpose but still near-optimal imaging protocol for objects emitting light incoherently, Pearce \emph{et al.}\ developed a method for estimating the complex degree of coherence (CDC) of the light emitted by the object \cite{Pearce}. The CDC can be used to reconstruct the intensity profile in  the source plane via a simple Fourier transform---the well-known Van Cittert-Zernike theorem---which for a thermal light source contains the complete information about the source distribution. This method was demonstrated experimentally by Howard \emph{et al.}\ \cite{Howard}. Technical challenges remain, as for the protocol the fields have to be recorded at different positions and superposed phase coherently during the detection, for which the detector positions must be known and kept fixed to within a fraction  of the wavelength of the light. 

Circumventing the need for phase coherence over large distances at optical frequencies is another option for beating the diffraction limit. Robert Hanbury Brown and Richard Q. Twiss famously developed optical intensity interferometry, using a pair of $6.5$~m telescopes acting as light detectors separated by up to $188$~m \cite{HBT_StellarII}. The method is not sensitive to the absolute phase of the fields at the location of the detectors, so the effective numerical aperture of the interferometer was on the order of $200$~m. In this way, Hanbury Brown and Twiss (HBT) successfully measured the diameter of Sirius A and many other  stellar diameters \cite{brown1956test,brown1974intensity}. 
Yet, a downside of HBT intensity interferometry is the low mode occupancy of thermal light, which means that recording a two-photon event in a single optical mode is exceedingly more rare as compared to a single-photon detection  as in amplitude interferometry. The average photon number at frequency $\omega$ in a single thermal mode is given by the degeneracy parameter $\delta = ({\rm e}^{\hbar\omega/k_{\rm B}T}-1)^{-1}$, where $k_{\rm B}$ is Boltzmann's constant and $T$ is the temperature of the light source \cite{Mandel95}. For light at a wavelength of 600~nm from the surface of a star at a  temperature of 5000~K we have $\delta \approx 8\times 10^{-3}$, whereas for near infrared wavelengths (900~nm) at the same temperature the degeneracy parameter is $\delta \approx 0.04$. This value is important for both interferometric methods, with the signal scaling linearly in $\delta$ for amplitude interferometry and quadratically in $\delta$ for intensity interferometry. 

However, $\delta$ is not the only relevant factor in extracting the information about the source separation in  both interferometric methods. Of equal importance is the numerical aperture of the interferometer which at optical frequencies can be larger by a factor of  $100$ and more for modern intensity interferometers  with respect to even todays's  largest amplitude interferometers. As it turns out, this advantage can more than compensate for the drawback of low degeneracy parameters in the optical domain, in particular considering very small stellar separations. 

Since $\delta$ is unfavourable for HBT interferometry, and because timing resolution and efficiencies of the detectors at the time of the original HBT experiments were restricted, intensity interferometry  was largely abandoned in the 1970s in favour of amplitude interferometry \cite{lawson2000principles}. 
Yet, modern detector technology has considerably advanced efficiencies, timing capabilities, data processing,  and synchronisation, and thus has enabled a new resurgence of HBT experiments with the goal of achieving improved resolution in astronomy. Recently, photon bunching, i.e, the temporal auto-correlation of photons,  was measured  with good signal to noise ratio for light from an artificial black body and the sun \cite{2014_Kurtsiefer, Tan2016}, from well-controlled laboratory sources \cite{Zmija2020, Matthews2018}, as well as from distant stars \cite{2016_NalettoFarisato_II4kmBaseline, 2017_GuerinKaiser_TemporalStellarII,2019_Matthews_VERITAS_II, Matthews2018a, Weiss2018, Klaucke2020}, and  first small baseline HBT experiments were carried out aiming to resolve true stars by use of spatial single photon \cite{2018_GuerinKaiser_SpatialStellarII, Rivet2020} or intensity \cite{2020_Acciarri_spatialI_MAGIC,2020_Kieda}  cross-correlations. For a recent review of stellar intensity interferometry, see Dravins \cite{Dravins16}. 

\begin{figure}[t!]
  \centering
  \includegraphics[width=.7\linewidth]{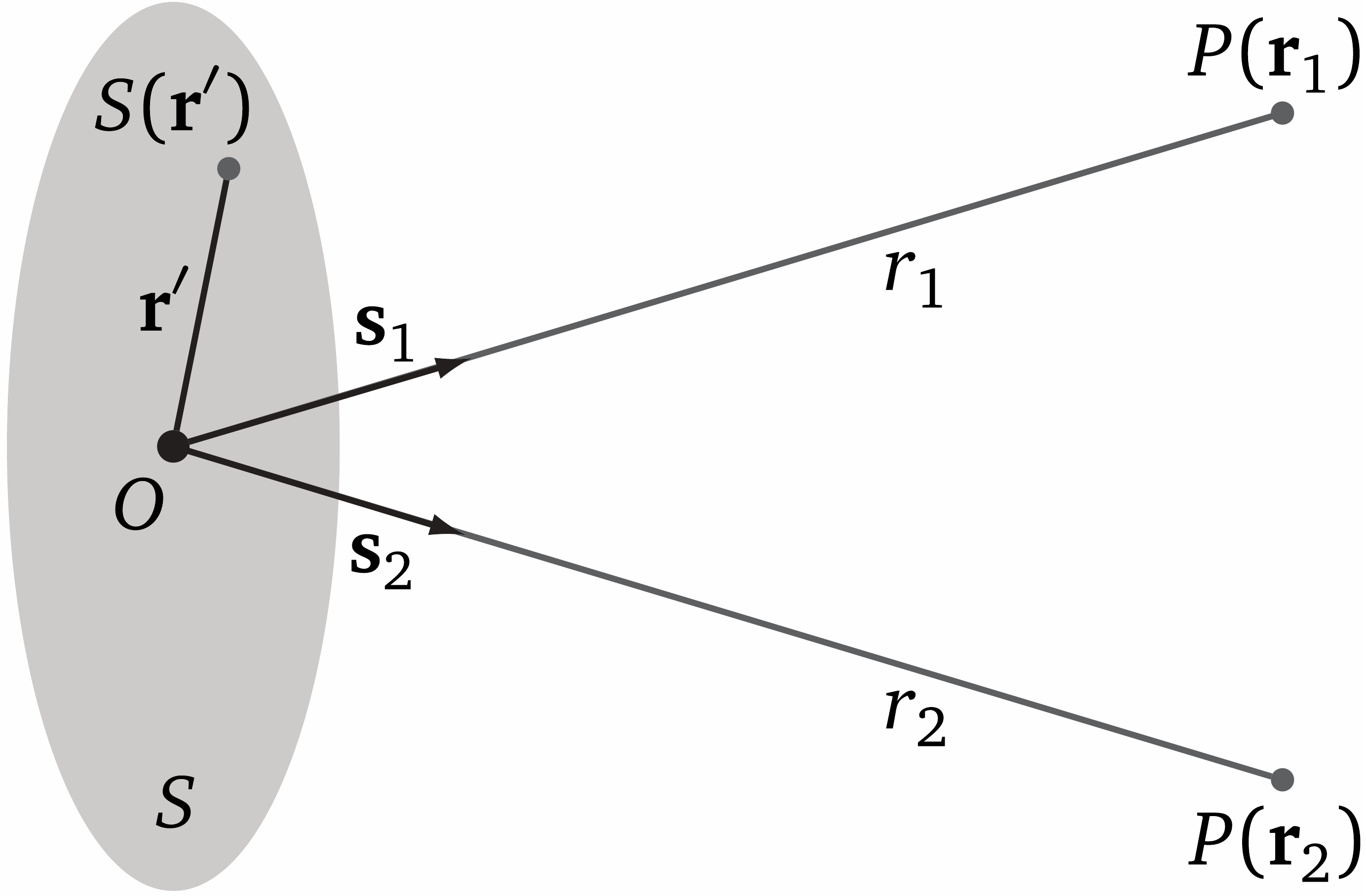}
  \caption{Complex degree of coherence in the far field of an incoherent source distribution $S(\mathbf{r}')$.  While the intensities $P(\mathbf{r}_1)$ and $P(\mathbf{r}_2)$  at each point $\mathbf{r}_1$ and $\mathbf{r}_2$ individually fluctuate randomly, the correlations of the field amplitudes at  $\mathbf{r}_1$ and $\mathbf{r}_2$ reveal the shape of $S(\mathbf{r}')$ since the complex degree of coherence $\gamma(\mathbf{r}_1,\mathbf{r}_2)$ depends on the geometrical details of $S(\mathbf{r}')$.}
  \label{fig:cdc}
\end{figure}

For thermal light sources like stars, one method of imaging the object of interest is knowing the CDC $\gamma(\mathbf{r}_1,\mathbf{r}_2)$, which measures the correlations between the electric field amplitudes at positions $\mathbf{r}_1$ and $\mathbf{r}_2$ in the far field of the object \cite{Mandel95}:
\begin{align}
 \gamma(\mathbf{r}_1,\mathbf{r}_2) = 
 \frac{\braket{E^*(\mathbf{r}_1)\, E(\mathbf{r}_2)}}
 {\sqrt{\braket{{E|(\mathbf{r}_1)}|^2} \braket{{|E(\mathbf{r}_2)}|^2} }}\, ,
\end{align}
where $\braket{E^*(\mathbf{r}_1)\, E(\mathbf{r}_2)} = G^{(1)}(\mathbf{r}_1,\mathbf{r}_2)$ is the first-order correlation function of the electric field at positions $\mathbf{r}_1$ and $\mathbf{r}_2$. 
The CDC is related to the source (intensity) distribution $S(\mathbf{r}')$ via the Van Cittert-Zernike theorem \cite{vanCittert34,Zernike38}
\begin{align}\label{eq:vfrhuw9}
 \gamma(\mathbf{r}_1,\mathbf{r}_2) = \frac{e^{i{k}({r}_2-{r}_1)}}{S_0} \int_S d\mathbf{r}'\, S(\mathbf{r}') e^{-ik(\mathbf{s}_2-\mathbf{s}_1)\cdot\mathbf{r}'} \, ,
\end{align}
where $S_0$ denotes the integrated source intensity and $k = 2\pi/\lambda$, with $\lambda$ the wavelength of the light (see Fig.~\ref{fig:cdc}). Further, $\mathbf{s}_1$ and $\mathbf{s}_2$ denote the unit vectors in the direction of $\mathbf{r}_1$ and $\mathbf{r}_2$. Hence, estimating  $\gamma(\mathbf{r}_1,\mathbf{r}_2)$ in the far field allows for the reconstruction of the source distribution $S(\mathbf{r}')$. All of the methods we are about to discuss in this paper rely on estimating $\gamma$. For the phase sensitive techniques based on amplitude interferometry, which we call $G^{(1)}$ methods, one has access to the complex quantity  $\gamma$. On the other hand, for intensity interferometry, which we call $G^{(2)}$  methods, we have  access only to the modulus of the CDC $|\gamma|$ via the normalized second-order correlation function $g^{(2)}(\mathbf{r}_1,\mathbf{r}_2)=1+|\gamma(\mathbf{r}_1,\mathbf{r}_2)|^2 $ \cite{Mandel95}. For simple objects such as resolving the diameter of a star, no involved phase information is needed and it is sufficient to measure the absolute value of the CDC. For more complex objects, however, the phase information of the CDC is essential since it is required for the reconstruction of the source intensity distribution via the Van Cittert-Zernike theorem in Eq.~(\ref{eq:vfrhuw9}). The $G^{(2)}$ methods can in principle overcome this difficulty by using well-established methods of phase recovery. There are several candidates, such as the Cauchy-Riemann approach \cite{Pittman95,Belenkii04}, Gerchberg-Saxon algorithms \cite{Fienup78,Fienup82,Kowalczyk90,Strekalov14}, and ptychographical intensity interferometry \cite{Wang18,Cao16}.

In this paper, we analyse quantitatively the relative strengths of the  different imaging methods described above using the (quantum) Fisher information. The latter is a well-established measure from estimation theory to quantify the amount of information  one can extract from the electromagnetic field with a given (optimal) measurement. We distinguish between (i) direct (Galilean) imaging using traditional phase interferometry by means of lenses and telescopes like the ELT, (ii) Tsang's small numerical aperture optimal imaging methodology [spatial mode demultiplexing (SPADE)], (iii) large aperture optimal interferometry like GRAVITY or CHARA, and (iv) imaging based on intensity interferometry such as HBT. The first three methods require access to the phase in the optical signal, including establishing a constant phase across the entire mirror of a telescope using active and adaptive optics, while the latter does not. 
We consider the theoretical limits to the resolution for the various interferometric methods taking into account the different achievable detector sizes as well as detector separations.
For an easier comparison of the methods, we will consider the separation  of two incoherent thermal point sources of potentially unequal brightness as our benchmark problem. Other important problems such as diameter estimation require {more refined estimation} techniques but do not lead to new bounds. 

The paper is organised as follows. In Sec.~\ref{sec:QFI} we present a short introduction to the (quantum) Fisher information. In Sec.~\ref{sec:G1} we review the various phase-coherent imaging methods, including traditional imaging and the new quantum imaging techniques, and examine some of their practical limitations. In Sec.~\ref{sec:G2} we discuss intensity interferometry for imaging, and compare it in Sec.~\ref{sec:comparison} to the amplitude interferometry techniques of Sec.~\ref{sec:G1}. In Sec.~\ref{sec:implementation}  we consider possible applications of intensity interferometry. We present our conclusions in Sec.~\ref{sec:discussion}.

\section{Tutorial: Fisher Information}\label{sec:QFI}\noindent
We provide a brief tutorial on the use of estimation theory in imaging and how to apply this to optimal---in a fundamental quantum mechanical sense---astronomical measurements. We can consider imaging problems as multi-parameter estimation problems. For example, the CDC $\gamma(\mathbf{r}_1,\mathbf{r}_2)$ for each pair of positions $\mathbf{r}_1$, $\mathbf{r}_2$ in the imaging plane is given by two parameters (e.g., the modulus $|\gamma|$ and its phase), and estimation theory can tell us what is the physical limit on the error in the parameter, given a particular measurement setup. This bound, the Cram\'er-Rao bound, is determined by the Fisher information (FI), i.e., the average information that is extracted in a measurement of the parameter. The \emph{quantum} Fisher information (QFI) is the Fisher information of the optimal quantum measurement, and is an intrinsic quantity of the quantum state (just like the entropy of a system is determined by its state and not how it is measured). 

\subsection{Fisher information}
\label{fisherinformation} 
\noindent When we measure a physical quantity, we are interested not only in the value of that quantity, but also the expected error in the measurement, typically in the form of the mean square error. Moreover, the quantity may not be measured directly, but rather is estimated indirectly based on the measurement of an observable that is related to it. As an example, consider the estimation of a phase difference $\theta$ between two arms of an interferometer. We cannot measure $\theta$ directly. Instead, we measure the photocurrent in the detectors at the output of the interferometer, and the measured currents allow us to infer a value of $\theta$. This is done via an estimator $\hat{\theta}(\mathbf{x})$, which is a function of the recorded data $\mathbf{x}$. Here, the data consists of the $N$ different measured photocurrents $\mathbf{x} = (M_1,\ldots,M_N)$. A nontrivial question is then how we determine the mean square error $\sigma_{\hat{\theta}}^2$ in $\theta$. It will be related to the variance $\sigma_M^2$ in the measured photocurrents according to the well-known error propagation formula
\begin{align} 
    \sigma_{\hat{\theta}}^2 = \frac{\sigma_M^2}{|d\<M\>/d\theta|^2}\, ,
\end{align}
where $\braket{M}$ is the expectation value of the photocurrents.
The quantity \smash{$\sigma_{\hat{\theta}}^2$} may be difficult to calculate directly due to the derivative $d\<M\>/d\theta$, and considerable effort has been spent to derive attainable lower bounds on \smash{$\sigma_{\hat{\theta}}^2$}. To this end, we define the Fisher information $F(\theta)$ as a functional of the probability distribution $p(x|\theta)$ over the possible measurement outcomes $x$ in the experiment
\begin{align} \label{eq:fi}
F(\theta) = \sum_x p(x|\theta) \left[ \frac{\partial \log{p(x|\theta)}}{\partial \theta} \right]^2 \, .
\end{align}
It is useful to think of the Fisher information as the average of the square of the so-called logarithmic derivative $\partial \log p(x|\theta)/\partial\theta$. The logarithm translates products of probabilities of independent events into a sum required to make the information additive: when we make two independent measurements of the system the information gained should double. Furthermore, the derivative with respect to the parameter $\theta$ enters because the more a probability distribution changes as we change $\theta$, the more information we expect to gain.

Given a model of the experiment, we can often determine $p(x|\theta)$ relatively easily, and by extension calculate $F(\theta)$. The mean square error $\sigma_{\hat{\theta}}^2$ is then bounded by the Fisher information and the number $N$ of statistically independent, identical repetitions of the experiment via
\begin{align} \label{eq:CRB}
\sigma^2_{\hat\theta} \geq [N F(\theta)]^{-1} \,.
\end{align}
This is the celebrated Cram\'er-Rao bound (CRB) on the precision with which we can estimate $\theta$. It assumes that the estimator is unbiased. In imaging and stellar interferometry applications the parameter $\theta$ may be the separation between two distant point sources, a stellar diameter, or the magnitude/phase of the CDC, among many others. The bound is generally attainable for a single parameter $\theta$ using the maximum likelihood estimator $\hat{\theta}_{\rm ML}(\mathbf{x})$ \cite{kay1993fundamentals}. If instead of a single parameter a parameter tuple $\bm{\theta}=(\theta_1,...,\theta_n)^\top$ is to be estimated, the logarithmic derivatives in Eq.~(\ref{eq:fi}) are taken with respect to two of the $n$ parameters. The Fisher information then becomes a matrix 
\begin{align} \label{eq:fimulti}
[F(\bm{\theta})]_{jk} = \sum_{x}  p(x|\bm{\theta}) \left[ \frac{\partial \log{p(x|\bm{\theta})}}{\partial \theta_j} \right]\left[ \frac{\partial \log{p(x|\bm{\theta})}}{\partial \theta_k} \right]\, ,
\end{align}
and Eq.~\eqref{eq:CRB} becomes a matrix bound on the covariance matrix of the parameters
\begin{align}
    \label{eq:multiCRB}
    \mathrm{Cov}(\bm{\theta}) \succeq \frac{1}{N} [F^{-1}(\bm{\theta})]\, ,
\end{align}
meaning that $\mathrm{Cov}(\bm{\theta})-F^{-1}(\bm{\theta})/N$ is a positive semidefinite matrix. 
This leads directly to a bound on the variances of the single parameters $\theta_j$ 
\begin{align}
\label{eq:clVarCRB}
    \mathrm{Var}(\theta_j) = [\mathrm{Cov}(\bm{\theta})]_{jj} \geq \frac{1}{N} [F^{-1}(\bm{\theta})]_{jj}\,.
\end{align}
Using, for example, a maximum likelihood estimator, this Cram\'er-Rao bound can be saturated.

As an example of how to apply estimation theory to imaging, first recall that Rayleigh's criterion \cite{rayleigh1879xxxi} establishes the minimum separation between two incoherent point sources required in order to resolve them. It arises from the diffraction of light through the finite aperture of the optical system, even as large as the  ELT. As a consequence, a point-like source will have a finite extension on the image plane, known as the Point Spread Function (PSF) \cite{pawley2010handbook,goodman2008introduction}. Two points that are closer together in the image plane than the width of the PSF will be difficult to resolve due to the substantial overlap of their images. The size of the PSF is of the order of the Rayleigh length $x_R = {1}/{k\mathrm{NA}}$, where $k$ is the wave number and $\mathrm{NA}$ is the numerical aperture that characterizes the optical system. 

For a Gaussian PSF with a width equal to the Rayleigh length, the image of a single point source at position $x_0$ (in the imaging plane) with unit intensity becomes
\begin{align}\label{eq:psf}
\psi_{x_0}(x) = \frac{1}{\sqrt[4]{2\pi x_R^2}} \exp \left[-\frac{(x-x_0)^2}{4x_R^2} \right]\, .
\end{align}
The corresponding normalised intensity profile is given by $|\psi_{x_0}(x)|^2$, which is equal to the probability distribution $p(x|x_0)$ of finding a photon at position $x$ in the imaging plane given that the PSF is centered around $x_0$. The parameter of interest is $x_0$, while the measurement data consists of values for $x$. The Fisher information for the measurement of the position $x_0$ is calculated from Eq.~(\ref{eq:fi}) as 
\begin{align} \label{eq:FIR}
F(x_0) = \frac{1}{4x_R^2}\,.
\end{align}
Substituting Eq.~(\ref{eq:FIR}) into 
Eq.~(\ref{eq:CRB}), we see that the variance of the PSF determines the precision in locating a single, point-like emitter, depending on the numerical aperture of the imaging system as expected. 

\begin{figure}[t]
\includegraphics[trim = 0cm 0cm 0cm 0cm, clip, width=.48\textwidth]{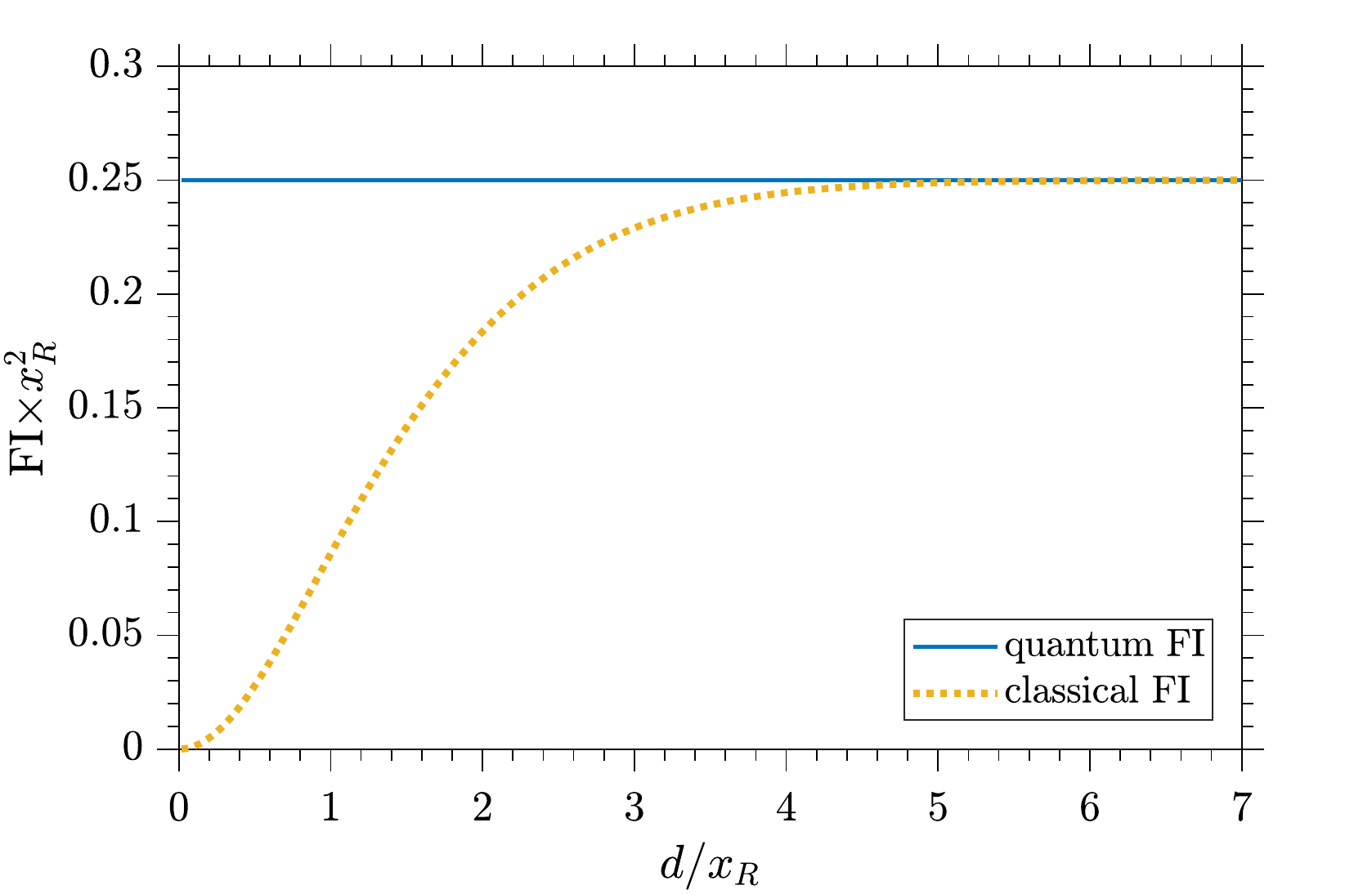} 
  \caption{\label{f:F21}
  Quantum Fisher information of estimating the separation $d$ of two equally bright thermal point sources [blue solid line, see Eq.~\eqref{eq:constQFI}] and classical Fisher information obtained via direct imaging [yellow dotted line, see App. \ref{sec:AppDirect}], as a function of $d$. The quantum Fisher information remains constant, whereas the classical Fisher information of direct imaging drops off and finally vanishes for  $d \rightarrow 0$.
  }
\end{figure}

Given a particular form of the PSF, one can also calculate the Fisher information for the {separation} $d$ {between two} equally bright sources \cite{PhysRevX.6.031033}. For a Gaussian point-spread function with variance $x_R^2$ it is calculated in Appendix \ref{sec:AppDirect}. The  result is shown  in Fig.~\ref{f:F21}. As can be seen from the figure,  for $d \gg x_R$ the Fisher information is a constant and equal to $\frac{1}{4} x_R^{-2}$, however, it rapidly approaches zero as $d$ drops below $x_R$. This phenomenon, dubbed the ``Rayleigh curse'', means that estimation becomes increasingly difficult when $d$ becomes smaller than $x_R$. Thus, estimation theory provides a rigorous foundation of the well-known Rayleigh criterion for image resolution. 

\subsection{Quantum Fisher information}
\label{sec:QFIsec}
\noindent While the Fisher information in classical estimation theory is a functional of the conditional probability distribution over the measurement outcomes given the value of the parameter $\theta$, in quantum estimation theory this becomes a functional of the quantum mechanical density operator describing the state of light. Two things should be noted: first, the density operator does not include any measurement outcomes, which instead are determined by the Hermitian observable that is measured, and second, the logarithmic derivative of an operator is not uniquely defined, giving rise to a certain amount of freedom in choosing the quantum Fisher information.

To define the quantum Fisher information, we replace the probability distribution $p(x|\theta)$ with the density operator $\rho(\theta)$, and we construct the so-called symmetric logarithmic derivative operator $L_\theta$ by solving {the Lyapunov} equation
\begin{align}\label{eq:lmvngjehut9wo}
    \partial_\theta \rho = \frac12 \left( L_\theta\rho + \rho L_\theta \right)\, ,
\end{align}
where $\partial_\theta \rho$ is the derivative of $\rho(\theta)$ with respect to $\theta$. The quantum Fisher information then becomes 
\begin{align}
    F_Q (\theta) = \tr\left[\rho L_\theta^2\right]\, .
\end{align}
The difficulty in calculating $F_Q$ is typically due to finding a closed form of the symmetric logarithmic derivative $L_\theta$. The quantum Cram\'er-Rao bound on the mean square error in $\theta$ then becomes 
\begin{align}
    \sigma_\theta^2 \geq [N F_Q(\theta)]^{-1}\, ,
\end{align}
where the optimal measurement is constructed on the eigenbasis of $L_\theta$. The bound is again attainable for a single parameter.

Imaging generally involves multiple parameters. For example, even the CDC for two positions in the imaging plane has two independent parameters, e.g., the magnitude and the phase of $\gamma$. We therefore extend the quantum Fisher information to the symmetric matrix
\begin{align}
    [F_Q(\bm\theta)]_{jk} = \tr\left[ \frac12 \rho \{L_j,L_k\}\right] ,
\end{align}
where the parameters are given by $\{\theta_j\}$, and each operator $L_j$ is defined according to Eq.~(\ref{eq:lmvngjehut9wo}) with respect to the corresponding parameter $\theta_j$. An analytic expression for the quantum Fisher information matrix (QFIM) reveals the dependency of estimation errors on different parameters, but is usually hard to compute and relies on finding a closed form of the symmetric logarithmic derivatives $L_j$. Usual approaches, relying on analytic matrix diagonalization, assume a representation of the density matrix in an orthogonal basis~\cite{Safranek2018} or even in its eigenbasis~\cite{liu2019quantum}. Recently, a nonorthognal-basis approach~\cite{Napoli2019,genoni2019non} was used to find a general analytic expression for the QFIM, which relies on matrix inversion via determining the general solution of the associated Lyapunov equations~\cite{Fiderer2021}. 

Once the QFIM is known, the quantum Cram\'er-Rao bound is a matrix bound on the covariance matrix over the parameters
\begin{align}
    \mathrm{Cov}(\bm\theta) \succeq \frac{1}{N} [F_Q^{-1}(\bm\theta)] \, .
\end{align}
For the variances of the different parameters, the bound implies 
\begin{align}
\label{eq:CRBVar}
    \mathrm{Var}(\theta_j) = [\mathrm{Cov}(\bm\theta)]_{jj} \geq \frac{1}{N} [F_Q^{-1}(\bm\theta)]_{jj}\, .
\end{align}
Unfortunately, the matrix bound is no longer tight in general. It is easy to see that this must be the case when the operators $L_j$ and $L_k$ are incompatible, i.e., do not commute. Then it is impossible to design a measurement observable that is optimal for both $\theta_j$ and $\theta_k$. This is a complication that we generally need to take into account, since the optimal measurement observables for, e.g., the phase and magnitude of the CDC are indeed incompatible. For a recent review on the quantum Fisher information and the fundamentals of quantum estimation theory, see Sidhu and Kok \cite{sidhu2020geometric}.

In their seminal work \cite{PhysRevX.6.031033}, Tsang, Nair and Lu showed that the Rayleigh limit for imaging can be avoided if we exploit more general measurement strategies.
They demonstrated that the quantum Fisher information for the separation between two weak thermal point sources with equal intensities is constant and independent of the value of the separation, i.e., 
$F_Q(d) = \frac14 x_R^{-2}$ (see Eq.~\eqref{eq:constQFI} and blue solid line in Fig.~\ref{f:F21}), whereas the classical Fisher information of direct imaging vanishes as the separation between the point sources approaches zero (see App. \ref{sec:AppDirect} and yellow dotted line in Fig.~\ref{f:F21}). This means that there is still a finite amount of information about the point source separation in the quantum state of light, even though the standard classical measurement does not have access to it. 

\begin{figure}
  \centering
  \includegraphics[width=.7\linewidth]{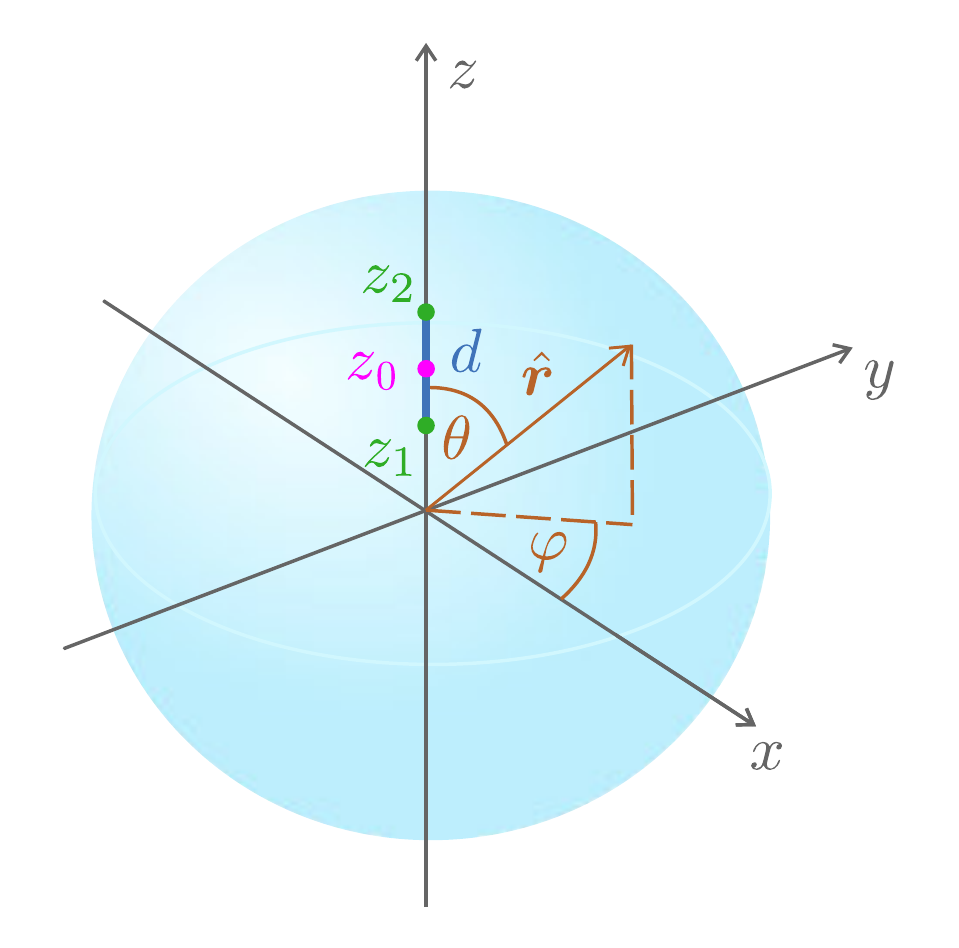}
  \caption{Setup of two thermal point sources placed along the $z$-axis at positions $z_1$ and $z_2$ around the centroid position $z_0$, separated by a distance $d$. A position in the far field is defined by the azimuthal angle $\varphi$ and the polar angle $\theta$ with respect to the coordinate origin.}
  \label{fig:Kugel1}
\end{figure}

In what follows, we  treat the optimal measurement in more detail.
We consider two thermal point sources emitting freely propagating photons into the far field.
The sources are placed at positions $\bm{z}_1$ and $\bm{z}_2$ along the $z$-axis around a centroid position $\bm{z}_0$ and separated by a distance $d$ (see Fig.~\ref{fig:Kugel1}). They can be described by a pair of complex amplitudes ${\bm A}=(A_-,A_+)^\top$ with respective photon numbers $N_{s,1} = \mathbb{E}_{\bm A}[|A_+|^2] = N_s q$ and $N_{s,2} = \mathbb{E}_{\bm A}[|A_-|^2] =N_s(1-q) $ \cite{Mandel95}, where $N_s$ is the total average photon number and $q \in [0,1]$ describes the relative strength of the sources. Here, $\mathbb{E}_{\bm A}$ denotes the expectation value with respect to ${\bm A}$. The density matrix of the light field produced by the two sources in the far field in the so-called Glauber-Sudarshan P-representation is~\cite{Mandel95}
\begin{equation}
\rho_d=\int_{\mathbb{C}^2}d^2A_+ d^2A_-\,P_{N_s}(\bm{A})\ket{\psi_{\bm{A},d}}\bra{\psi_{\bm{A},d}}\,,
\label{eq:TLS1.1}
\end{equation}
where the joint probability distribution is given by the product of the individual probability distributions as
\begin{equation}
P_{N_s}(\bm{A})=P_{N_{s,1}}(A_+)P_{N_{s,2}}(A_-)
\end{equation}
with
\begin{equation}
P_{N_{s,1/2}}(A_\pm) = \left(\frac{1}{\pi N_{s,1/2}}\right)\exp\left(-\frac{\left| A_\pm\right|^2}{N_{s,1/2}}\right) \, .
\label{eq:GlauberP}
\end{equation}
In Eq.~\eqref{eq:TLS1.1},  $\ket{\psi_{\bm{A},d}}$ represents a coherent state in the far field with mean number of photons $N_s$, which is an eigenstate of the positive frequency field operator $E(\mathbf{r})$
\begin{equation}
E(\mathbf{r})\ket{\psi_{\bm{A},d}}=\psi_{\bm{A},d}(\mathbf{r})\ket{\psi_{\bm{A},d}}\,,
\end{equation}
with eigenvalue
\begin{equation}
\psi_{\bm{A},d}(\mathbf{r})=A_+\psi(\mathbf{r}-\bm{z}_2)+A_-\psi(\mathbf{r}-\bm{z}_1) \, ,
\end{equation}
where $\bm{z}_{1,2} = \bm{z}_0 \mp \bm{d}/2$, $\bm{d}=(0,0,d)^\top$, and $\psi(\mathbf{r})$ denotes the photonic wave function in the far field.

An important quantity is the overlap between the wave functions for a given translation $\bm{a}$
\begin{equation}
\eta(\bm{a})\ = \int d^2r\,\psi^*(\mathbf{r})\psi(\mathbf{r}-\bm{a}) \, .
\end{equation}
In general, $\eta(\bm{a})$ is non-zero, which means that the wave-functions coming from the two sources are non-orthogonal spatially, which is what makes estimation based on photon counting on a screen difficult. 

In the case of thermal light sources, the average photon number per mode is usually very small, which allows us to truncate the photonic state after the one-photon state. Under the assumption that the overlap function is real and symmetric, i.e., 
\begin{equation}
\eta(\bm{a})=\eta^*(\bm{a})=\eta(-\bm{a})\, ,
\label{eq:OverlapCond}
\end{equation}
the quantum Fisher information matrix for estimating the centroid $z_0$, the separation $d$ and the relative strength $q$ can be calculated as~\cite{SanchezSoto2017}
\begin{equation}
F_Q=4\begin{pmatrix}
p^2+4q(1-q)\wp^2 & (q-\frac12)p^2 & -iw\wp\\
(q-\frac12)p^2 & \frac14 p^2 & 0\\
-iw\wp & 0 & \frac{1-w^2}{4q(1-q)}
\end{pmatrix}\,.
\label{eq:QFIM}
\end{equation}
Here, the different quantities entering the quantum Fisher information matrix are
\begin{align}\label{eq:ueghordf}
    w & =  \braket{\psi|\exp(i d P)|\psi}  = \eta(d) = \exp\left(-\frac{d^2}{8 x_R^2}\right) \, ,\cr
    p^2 & = \braket{\psi|P^2|\psi} = \frac{1}{4 x_R^2}\, ,\\
    \wp & = \braket{\psi|\exp(idP)P|\psi}= \frac{i d }{4 x_R^2}\exp\left(-\frac{d^2}{8 x_R^2}\right) \, ,\nonumber
\end{align}
where $\ket{\psi}=\int dx\, \psi(x)\ket{x}$,  $P=-i\partial_x$, and the analytical values were calculated using the Gaussian PSF of Eq.~\eqref{eq:psf} (with $x$ replaced by $z$). One can see that in case of $q\neq \frac12$, i.e., sources of different strength, all three parameters $z_0$, $d$ and $q$ become intertwined, such that it is a non-trivial task to estimate either of the three parameters. 

\begin{figure}[t]
\includegraphics[trim = 0cm 0cm 0cm 0cm, clip, width=.5\textwidth]{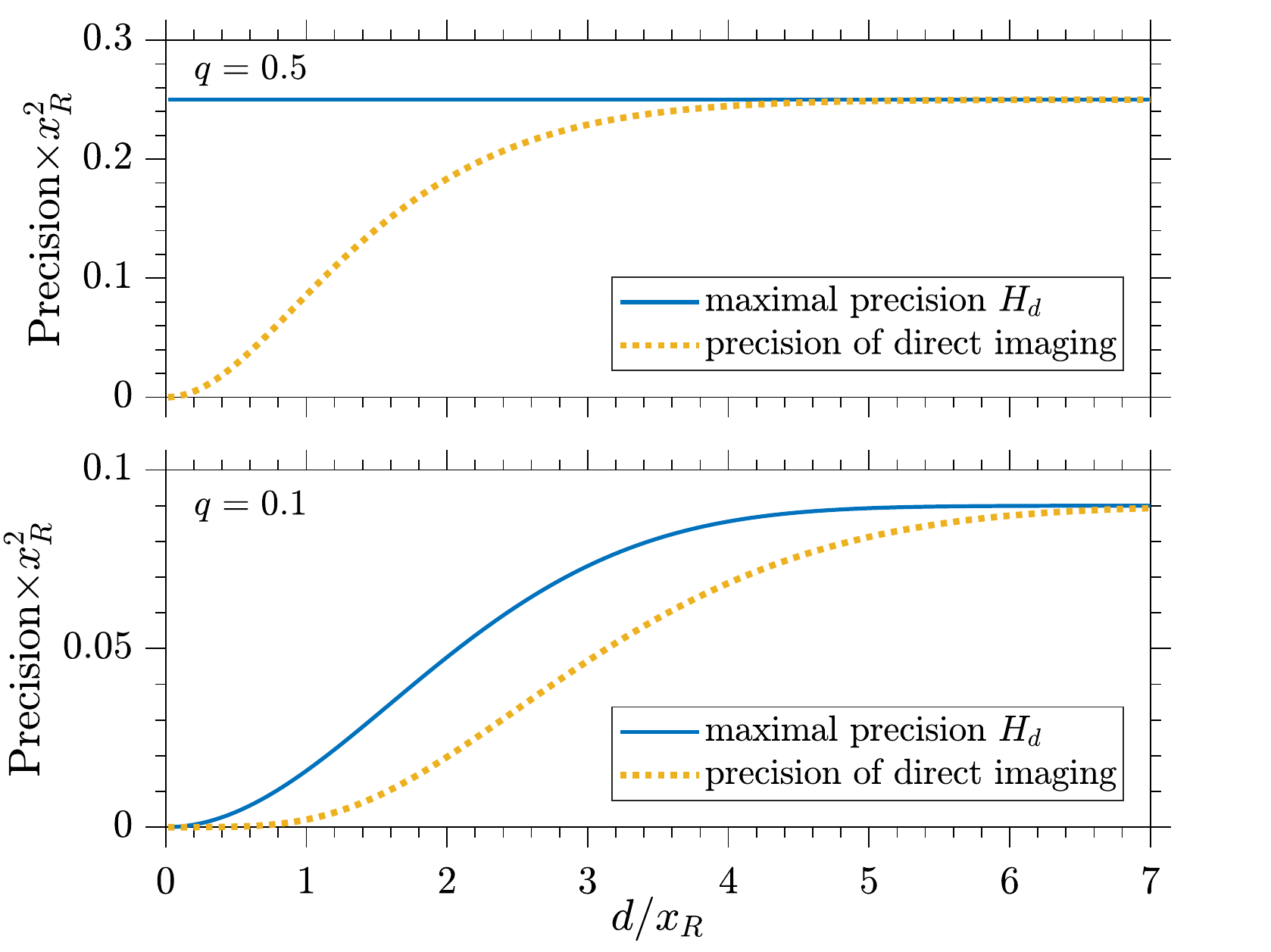} 
  \caption{\label{f:F2}
  Maximal precision of the separation $d$ of two thermal point sources given by $H_d$ in Eq.~\eqref{eq:prec_gauss} (blue solid line), and   precision obtained via direct imaging (yellow dotted line, see App. \ref{sec:AppDirect}) as a function of $d$. For the special case of equally bright sources ($q=\frac12$), the maximal precision remains constant, whereas direct imaging is not capable of achieving this precision in the limit $d\rightarrow 0$. However, for unequally bright sources ($q\neq \frac12$) the "Rayleigh curse" is unavoidable, meaning that even the maximal precision $H_d$ drops off to zero for $d\rightarrow 0$,  although it remains always higher than the precision obtained via direct imaging.
  }
\end{figure}

However, the precision in a particular parameter can be calculated by taking the inverse of the diagonal elements of the inverse quantum Fisher information matrix. In this way, the precision about the separation of the sources $d$ is  found to be~\cite{SanchezSoto2017} 
\begin{equation}
    H_d = p^2\mathscr{D}^2\frac{\wp^2+p^2(1-w^2)}{\mathscr{D}^2\wp^2+p^2(1-w^2)}\,,
    \label{eq:InfQ}
\end{equation}
where $\mathscr{D}^2\equiv 4q(1-q)$. 

In the special case of equally bright sources ($q=\frac12$), one finds from Eq.~\eqref{eq:InfQ} that
\begin{equation}
\label{eq:constQFI}
    H_d = p^2\,,
\end{equation}
i.e., a constant with respect to the separation $d$ of the sources. This is the famous result found by Tsang, Nair and Lu~\cite{PhysRevX.6.031033}. 
By contrast, for $q \neq \frac12$, the precision $H_d$ is not a constant as a function of the separation $d$. Indeed, in the case of a Gaussian PSF, by substituting Eq.~(\ref{eq:ueghordf}) into Eq.~(\ref{eq:InfQ}), $H_d$ becomes
\begin{align}
\label{eq:prec_gauss}
 H_d = \frac{k^2 \mathrm{NA}^2 (1-q)q\left[ \frac14 d^2 k^2 \mathrm{NA}^2 - \exp( \frac14 d^2 k^2 \mathrm{NA}^2) + 1 \right]}{d^2 k^2 \mathrm{NA}^2 (1-q)q - \exp( \frac14 d^2 k^2 \mathrm{NA}^2) + 1 } ,
\end{align}
where we have explicitly substituted $x_R = {1}/{k\mathrm{NA}}$ to highlight the dependence on the numerical aperture $\mathrm{NA}$. For large separations $d \gg x_R = 1/k\mathrm{NA}$, i.e., $d k \mathrm{NA}\gg 1$, the exponentials in the denominator and the numerator dominate and cancel each other, such that 
\begin{align}
\label{eq:Hdlarged}
    H_d \approx k^2\mathrm{NA}^2(1-q)q=\text{constant.}
\end{align}
In the case of the other extreme, i.e., $d\rightarrow 0$, we can expand Eq.~\eqref{eq:prec_gauss} in a Taylor series, which gives to lowest order
\begin{align}
    H_d \approx \frac{k^4 \mathrm{NA^4} (1-q)q}{8(1-2q)^2}d^2\,.
\end{align}
In this limit, the precision drops to zero quadratically in $d$. In between these limiting cases we can identify the point at which $H_d$ is reduced by 3~dB:
\begin{align}
 d \simeq \frac{2\sqrt{2-8q+8q^2}}{k \mathrm{NA}} = 2\sqrt{2-8q+8q^2}\;x_R\, ,
\end{align}
which is approximately $2.26\, x_R$ for $q=0.1$. Thus, the numerical aperture crucially determines the starting point of the drop-off, i.e., for larger numerical apertures the drop-off starts only at smaller separations $d$. This feature is the essential aspect in the comparison of amplitude versus intensity interferometry accomplished in Sec.~\ref{sec:comparison}.

In Fig.~\ref{f:F2} we show the precision in Eq.~\eqref{eq:prec_gauss} for $q=\frac12$ and $q\neq \frac12$, together with the precision obtained via the non-optimal measurement of direct imaging (see App. \ref{sec:AppDirect}). As can be seen in Fig.~\ref{f:F2}, the precision achievable by direct imaging remains always smaller than the one obtained via optimal imaging, the former only approaching the optimal precision for $d\gg x_R$. Further, for $q=\frac12$, the maximal precision is a constant with respect to the separation of the two sources [see Eq.~\eqref{eq:constQFI}], whereas for $q\neq \frac12$ the ``Rayleigh curse'' is unavoidable, even for the optimal case.

In this section we have given a broad outline for how to use the quantum Fisher information in imaging situations, and applied the theory to the estimation of the separation between two distant thermal point sources. The quantum Cram\'er-Rao bound places a limit on the precision with which such a parameter can be measured, but the bound does not immediately tell us how this precision can be achieved. For that we need to either calculate the symmetric logarithmic derivative operator $L_d$, or show that a given measurement saturates the quantum Cram\'er-Rao bound. This will be the subject of the next two sections.

\section{amplitude interferometry}\label{sec:G1}\noindent
Next, we consider astronomical imaging methods that rely on amplitude interferometry, i.e., 
measuring the $G^{(1)}$ correlation function in the far field. In other words, these are phase-sensitive measurements. We have already  considered direct (Galilean) imaging using traditional phase interferometry by means of lenses and telescopes like the ELT in Sec.~\ref{fisherinformation} (see also App. \ref{sec:AppDirect}). 
For a one-dimensional problem, the PSF of a single point source is proportional to $\text{sinc}^2 x$ which can be approximated by the Gaussian function given in Eq.~\eqref{eq:psf}. In that case, we showed the Fisher information for the separation $d$ between two thermal point sources for two different values of the relative brightness $q$ in  Fig.~\ref{f:F2}.
In the next two sub-sections, we consider the method employed by Nair, Tsang and Lu, called SPADE, providing an optimal measurement of the separation $d$ between two thermal point sources (Sec.~\ref{subsec:G1B}), and  investigate two-mode interferometry that may be easier to implement, yet achieves the same resolution as SPADE (Sec.~\ref{subsec:G1C}).

\begin{figure}[t!]
  \centering
  \includegraphics[trim = 0cm 0 0 0, clip, width=0.8\linewidth]{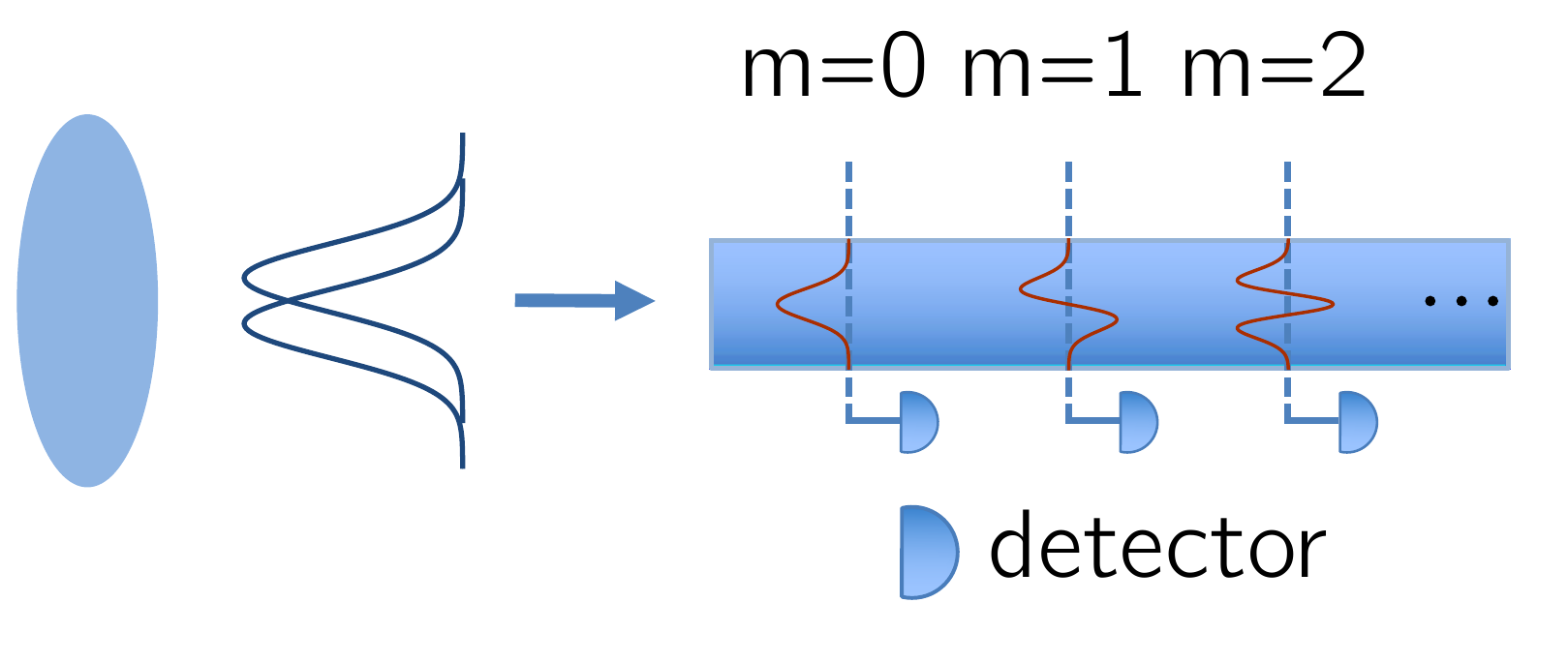}
  \caption{Illustration of the setup for SPADE:
  A multimode waveguide performs Hermite-Gaussian spatial mode sorting; photons of different Hermite-Gaussian modes are separated out and sent to different photon counters. The centroid of the two sources is assumed to be  aligned with the center of the waveguide.}
  \label{fig:spade}
\end{figure}

\subsection{Optimal amplitude interferometry  - SPADE}\label{subsec:G1B}

\begin{figure}[t]
  \centering
  \includegraphics[trim = 0cm 0 0 0, clip, width=1\linewidth]{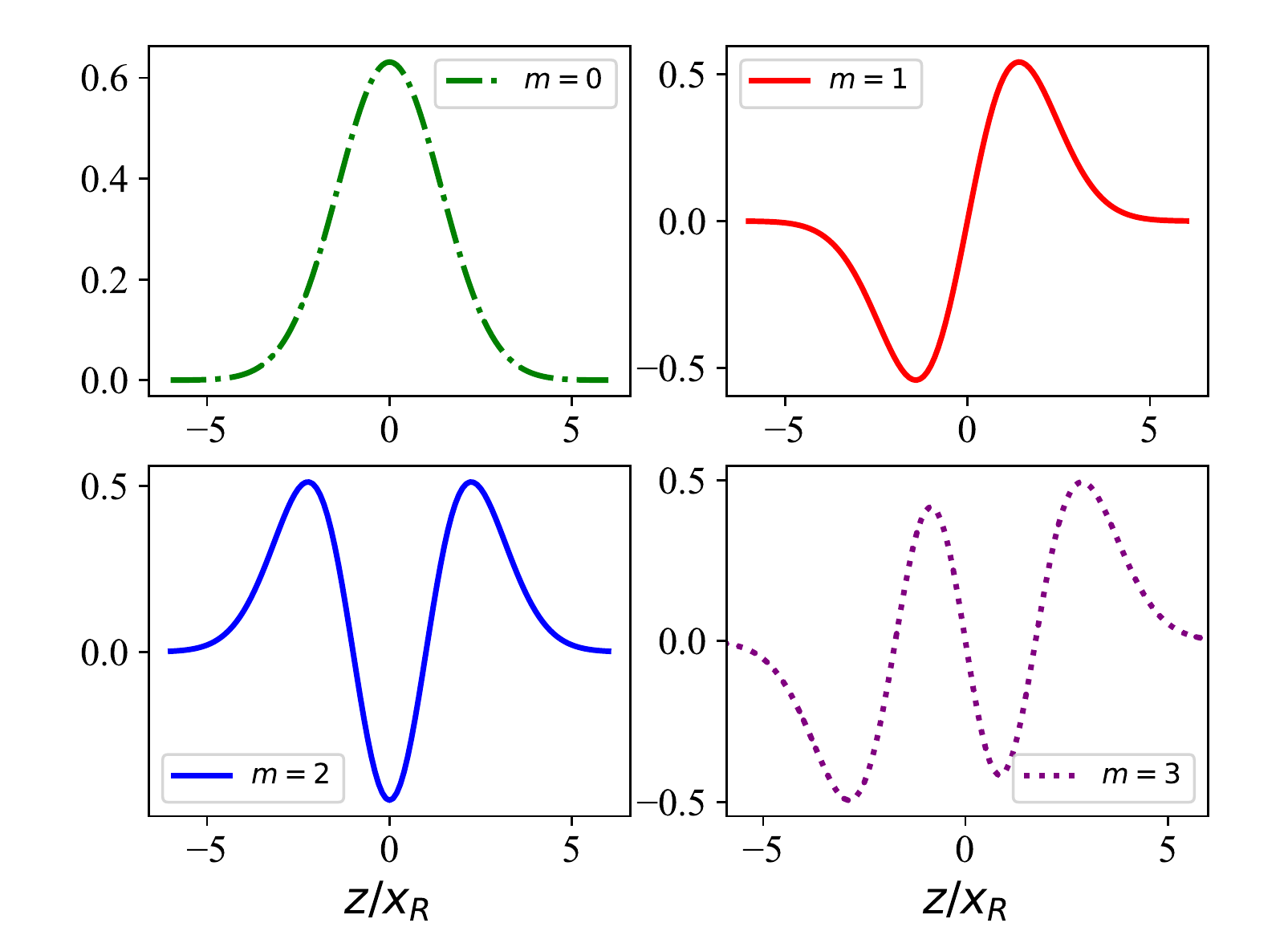}
  \caption{Example of Hermite-Gaussian mode profiles, for $m=0,1,2,3$. The incoming signal excites different higher-order Hermite-Gaussian modes, providing information on the separation. 
  }
  \label{fig:HG} 
\end{figure}

\noindent
Assuming that the PSF is Gaussian, we can devise an optimal measurement for estimating the angular separation between two sources. Rather than measuring the distribution of the photons on a screen, we decompose the light into Hermite-Gaussian spatial modes. To achieve this, our imaging system is assumed to consist of a multi-mode wave guide aligned such that the centroid of the two sources lies on the central axis of the wave guide. 
The signal is then demultiplexed so that photons of different Hermite-Gaussian modes in the wave guide are sent to different detectors (see Fig.~\ref{fig:spade}). The Hermite-Gaussian spatial modes can be formally denoted by the basis states $\{\ket{\Pi_m}; m = 0,1,2,... \}$, with 
\begin{align}
\ket{\Pi_m} &= \int_{-\infty} ^\infty dz ~\Pi_m(z) \ket{z}, \qquad m=0,1,2,... \nn
\Pi_m(z) &= \left(\frac{1}{2\pi x_R^2}\right)^{\frac{1}{4}} \frac{1}{\sqrt{2^m m!}} 
H_m\left(\frac{z}{\sqrt2 x_R} \right)
\exp\left(-\frac{z^2}{4x_R^2} \right) \, ,
\end{align}
\noindent where $H_m$ is the Hermite polynomial of order $m$, and $z$ is the transverse position in the imaging plane \cite{yariv1989quantum}. The mode profiles for the first four modes are shown in Fig.~\ref{fig:HG}. The probability that the photon is measured in mode $m$ is given by the Born rule $\tr(\rho \ket{\Pi_m}\bra{\Pi_m})$, where $\rho$ is the state of light in the wave guide. 

Next, we calculate the Fisher information achieved by SPADE for equally bright sources and show that it is equal to the quantum Fisher information.
For a Gaussian PSF the probability that a photon is detected in a particular mode $H_m$ is \cite{PhysRevX.6.031033}
\begin{align} \label{eq:iq}
 p(m) &=  \exp(-Q)\frac{Q^m}{m!}\, ,
\end{align}
with
\begin{align} 
Q &\equiv \frac{\zeta^2}{16 x_R^2}\, ,
\end{align}
where we assume that the two distant point sources produce PSFs in the imaging plane centered at $z=\pm \zeta/2 $, i.e.,  $\zeta $ is the separation of the sources on the image screen.
Using Eq.~\eqref{eq:fi}, the Fisher information for the parameter $\zeta$ is calculated as 
\begin{align}
F({\zeta}) \approx N_s \sum_{m=0}^\infty p (m) \left[ \frac{\partial}{\partial \zeta} \ln p (m) \right]^2 \approx \frac{N_s}{4 x_R^2}\, ,
\end{align}
where $N_s$ is the mean photon number of the sources.
This is the classical Fisher information achieved by SPADE. Note that in the limit $N_s \ll 1$ and for $q=\frac12$, it follows from Eq.~\eqref{eq:prec_gauss} that the QFI is also equal to $N_s/(4 x_R^2)$. Hence, the Fisher information of SPADE achieves the theoretical maximum given by the quantum Fisher information, i.e.,  SPADE corresponds to an optimal measurement of the separation between the two sources.

Intuitively, SPADE works for the following reason: given that the two distant point sources produce PSFs in the imaging plane centered at $z=\pm \zeta/2 $, the wave function for small $\zeta$ can be approximated by
 \begin{align}
     \psi\left(z\pm \frac{\zeta}{2}\right) \approx \psi(z) \pm \frac{\zeta}{2}          \frac{\partial \psi(z)}{\partial z}\,.
 \end{align}
The function $\psi(z)$ is proportional to the $m=0$ Hermite-Gaussian mode, while  ${\partial \psi(z)}/{\partial z}$ is proportional to the $m=1$ Hermite-Gaussian mode. Since these are orthogonal, they can be distinguished. The fundamental  $m=0$ spatial mode is insensitive to the parameter $\zeta$. The $m=1$ mode, however, \textit{does} contain information about $\zeta$, since $\Pi_1(z)$ is proportional to ${\partial \psi(z)}/{\partial z}$. The relative probabilities $p(m)$ for finding photons in the different Hermite-Gaussian modes provide the optimal estimator for $\zeta$, which  can be related directly to the angular separation $\theta = d/r$, the parameter of interest.
Here, $d$ is the distance between the sources, and $r$ is the distance from the centroid to the imaging system.

Implementing  SPADE  is difficult, as different spatial modes need to be separated into different channels before detection. Several physical devices has been proposed \cite{tsang2019resolving}, and a series of experiments \cite{paur2016achieving,tang2016fault,yang2016far,PhysRevLett.118.070801,PhysRevLett.121.090501,PhysRevLett.121.250503,paur2018tempering,hassett2018sub,zhou2019quantum} has confirmed the feasibility of SPADE-type measurements. The most accessible implementation remains an open question.

\subsection{Optimal two-mode interferometry}\label{subsec:G1C}
\noindent
SPADE effectively uses only the lowest two Hermite-Gaussian modes $m=0$ and $m=1$. This suggests that more generic two-mode interferometry may also provide optimal measurements of the source separation. The advantage is that two-mode interferometry is a mature technology. In this section we discuss the relative merits of two-mode interferometry for quantum imaging. 

\begin{figure}[t!]
  \centering
  \includegraphics[width=1.0\linewidth]{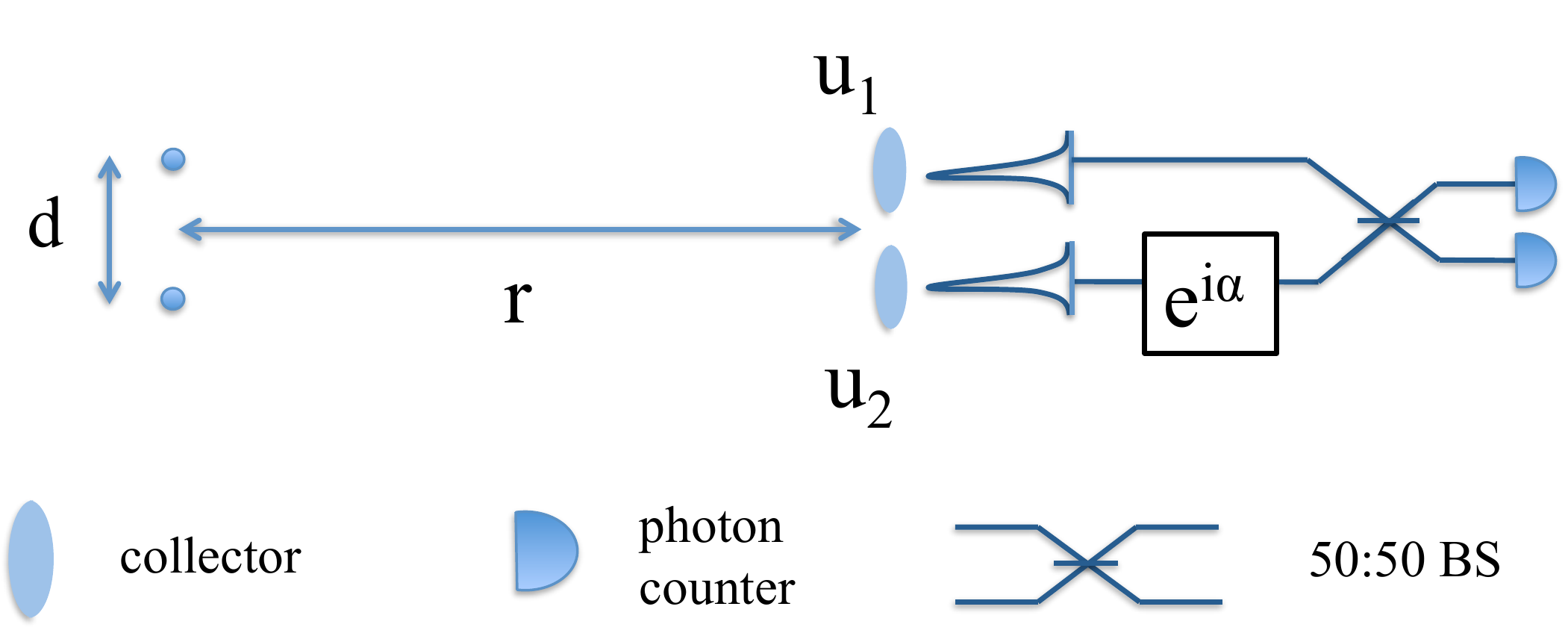}
  \caption{Illustration of the setup for two-mode interferometry. Light is collected at positions $u_1$ and $u_2$ and fed into single-mode wave guides. A variable phase $\alpha$ may be applied to one of the modes. A beam splitter mixes the incoming modes before detection in photon counters.}
  \label{fig:two_mode}
\end{figure}

Let us consider two light-collecting fibers (collectors) that are separated by a distance $|u_1-u_2|$. Assuming two equally bright incoherent point sources separated by a distance $d$, the QFI for the angular separation $\theta$ can be calculated as \cite{PhysRevLett.124.080503}
\begin{align}
F_Q(\theta) = \frac{k^2}{4} \left( u_1 - u_2\right)^2.\label{eq:gnjehitu39w}
\end{align}
This expression is independent of $\theta$, i.e., the separation $d$ of the point sources, and therefore the measurement of the separation is in principle not limited by diffraction. 
A simple experimental scheme that can achieve this precision is shown in Fig.~\ref{fig:two_mode}. 

In what follows, we calculate the Fisher information for the setup of Fig.~\ref{fig:two_mode} and show as before that its classical Fisher information is identical to the quantum Fisher information of Eq.~(\ref{eq:gnjehitu39w}). Therefore, like SPADE, the setup in Fig.~\ref{fig:two_mode} is an optimal measurement of the source separation.

We can write the state of a single photon originating from the two sources as
\begin{align}
\label{eq:densitytm}
\rho \approx \frac{1}{2}\ket{\Psi_1}\bra{\Psi_1} + \frac12\ket{\Psi_2}\bra{\Psi_2}\, .
\end{align}
Eq.~(\ref{eq:densitytm}) corresponds to an incoherent mixture of the photon originating from source 1 (denoted by $\Ket{\Psi_1}$) and source 2 (denoted by $\Ket{\Psi_2}$). As the photons propagate towards the collectors at $u_1$ and $u_2$, the single photon states evolve to a very good approximation into (see Fig.~\ref{fig:two_mode})
\begin{align}
\label{eq:int_states}
\ket{\Psi_1} &\approx \frac{1}{\sqrt2} \left( \ket{\psi_{u_1}} +   e^{i  \phi + \alpha}\ket{\psi_{u_2}} \right),\nn
\ket{\Psi_2} & \approx\frac{1}{\sqrt2} \left(\ket{\psi_{u_1}}  +   e^{-i  \phi + \alpha}\ket{\psi_{u_2}} \right)\, ,
\end{align}
where $\ket{\psi_{u_1}}$ is the state of a photon collected in $u_1$, $\ket{\psi_{u_2}}$ is the state of a photon collected in $u_2$, and the phase $\phi$ is given by 
\begin{align}
\phi \approx \frac12 (u_1-u_2) k \theta\,,
\end{align}
to an excellent approximation. Setting the adjustable phase $\alpha=0$, the collected light is interfered at a 50:50 beam splitter, and the probabilities of finding the photon in detector 1 and detector 2, respectively, are given by
\begin{align}
p_1=\frac{1}{2} (1+\cos \phi ) \quad\text{and}\quad
p_2=\frac{1}{2} (1-\cos \phi )\, .
\label{eq:2_det_ps}
\end{align}
Substituting these probabilities into the expression for the Fisher information in Eq.~\eqref{eq:fi} yields
\begin{align}
F({\theta}) &= \sum_{i = 1,2}p_i  \left[ \frac{\partial}{\partial \theta} \ln p_i \right]^2
=\frac{k^2}{4} (u_1 - u_2)^2\, ,
\end{align}
which is independent of $\theta$ and coincides with the quantum Fisher information in Eq.~(\ref{eq:gnjehitu39w}). 
We note that for two-mode interferometry a similar phenomenon occurs as with SPADE, i.e., equal brightness of the two sources leads to a constant quantum Fisher information, whereas unequal brightness gives rise to a drop-off of the quantum Fisher information as the separation between the sources approaches zero \cite{PhysRevLett.124.080503}.

The two-mode interferometric scheme has essentially the same resolution as SPADE. However, it removes the need for building a large system of lenses and mirrors for the telescope, as well as the experimental complexity for the SPADE measurement. Yet, the former comes at the cost of a reduced photon count rate by a factor of the ratio of the collecting areas. Also, in order to extract the angular separation from the measurements, the entire interferometer  needs to be phase-stabilised to the same accuracy as any other $G^{(1)}$ measurement, i.e., to a fraction of a wavelength for the duration of the measurement. In astronomy, such optimal quantum telescopes have been implemented, e.g., in the GRAVITY interferometer as well as in CHARA.

\section{Intensity interferometry}\label{sec:G2}
\noindent
Next, we investigate the precision obtained via $G^{(2)}$ intensity interferometry, which requires correlating photons recorded at two detectors. Since the numerical aperture is defined by the distance between the two detectors, e.g., two telescopes, 
and since the correlations are generated only  electronically, sometimes even only via post-processing, this setup allows for substantially larger numerical apertures than those achievable with the $G^{(1)}$ methods discussed in the  previous section. Moreover, the telescopes can be simple in design as the photons have to be recorded only  within their coherence time (or several coherence times, depending on the time resolution of the detectors), given, e.g., by the bandpass of the filter used in the setup. This is in  contrast to  amplitude interferometry where a phase-sensitive measurement of the electromagnetic field is needed, more likely to be affected by small perturbations, e.g., optical imperfections of the detectors or turbulences in the atmosphere. Eliminating such perturbations requires an elaborate design of the telescopes such as adaptive or active optics, limiting the  numerical aperture of amplitude interferometry to typically $< 500$ m. Intensity interferometry on the other hand is not disturbed by detector imperfections or atmospheric turbulences. Moreover, the  signal only depends on the source separation $d$ and the relative strength $q$ of the sources, but is insensitive to the centroid position $z_0$ of the light sources. In total, intensity correlation measurements appear simpler to implement while providing larger numerical apertures at potentially lower costs. The question is whether these advantages can overcome the reduced signal strength related to the small degeneracy parameter $\delta$.

In what follows we investigate the precision of $G^{(2)}$ intensity correlation measurements using a lower bound on the Fisher information. Considering realistic numerical apertures and photon fluxes, we  will see that $G^{(2)}$ measurements indeed scale favorably in certain regions of the angular separation of the sources, in particular at  very  small separations,  even with respect to the quantum Fisher information of optimal $G^{(1)}$ measurements.

In general, for the state given in Eq.~\eqref{eq:TLS1.1}, consisting of an infinite sum over different photon numbers, the calculation of the Fisher information for a particular measurement is challenging. We will thus compute only a general lower bound on the Fisher information to estimate an unknown parameter array and compare it to the various $G^{(1)}$ methods discussed in the former section. However, for a fair comparison, we will consider different numerical apertures for $G^{(2)}$ and $G^{(1)}$ interferometry, i.e., we will allow for the reasons described above for the $G^{(2)}$ measurements much larger  numerical apertures than for the  $G^{(1)}$ measurements. 

The intensity correlation function $G^{(2)}(\mathbf{r}_1,\mathbf{r}_2)$ at positions $\mathbf{r}_1$ and $\mathbf{r}_2$ is defined as
\begin{align}\label{eq:G2}
G^{(2)}(\mathbf{r}_1,\mathbf{r}_2) & = \braket{ E^\dagger(\mathbf{r}_1)E^\dagger(\mathbf{r}_2)E(\mathbf{r}_2)E(\mathbf{r}_1)} \cr
 & =\braket{I(\mathbf{r}_2) I(\mathbf{r}_1)} - \delta(\mathbf{r}_1 - \mathbf{r}_2)\braket{ I(\mathbf{r}_1) } ,
\end{align} 
where $E^\dagger(\mathbf{r})$ and $E(\mathbf{r})$ are the negative and positive frequency parts of the electric field operator at position $\mathbf{r}$. Since the operators  $E^\dagger(\mathbf{r})$ and $E(\mathbf{r})$  do not commute, an extra term appears in Eq.~(\ref{eq:G2}) for the case that $\mathbf{r}_1=\mathbf{r}_2$. 

We suppose the detectors used for the $G^{(2)}$ measurements to have an area of $A = 10\,\mathrm{m}\times 10\,\mathrm{m}$. This is a reasonable size taking into account that the telescopes are only photon collecting detectors  which neither need  adaptive nor active optics. 
The numerical aperture spanned by one detector is approximately $\sqrt{A}/r$, with $r$ the distance from the stellar sources to the detector on Earth. However, the numerical aperture of the entire $G^{(2)}$ interferometer is given by the distance between the outermost detectors. In order to exploit multiple correlations at different detector separations, we will consider an ensemble of altogether $100$  detectors, located at $D$ different positions. This number of telescopes equals  the number of detectors used for  the system presently predominantly  investigated for an implementation of  $G^{(2)}$ intensity interferometry in astronomy, the  Cherenkov Telescope Array (CTA) \cite{CTASouthObservatory2021,2019_Kieda, 2012_Dravins}. The sizes of the telescopes used in CTA (between $4$\;m and $23$\;m in diameter) span a similar range  as assumed in our calculation and cover approximately the same total area. 

\begin{figure}[t!]
 \centering
 \includegraphics[width=.5\textwidth]{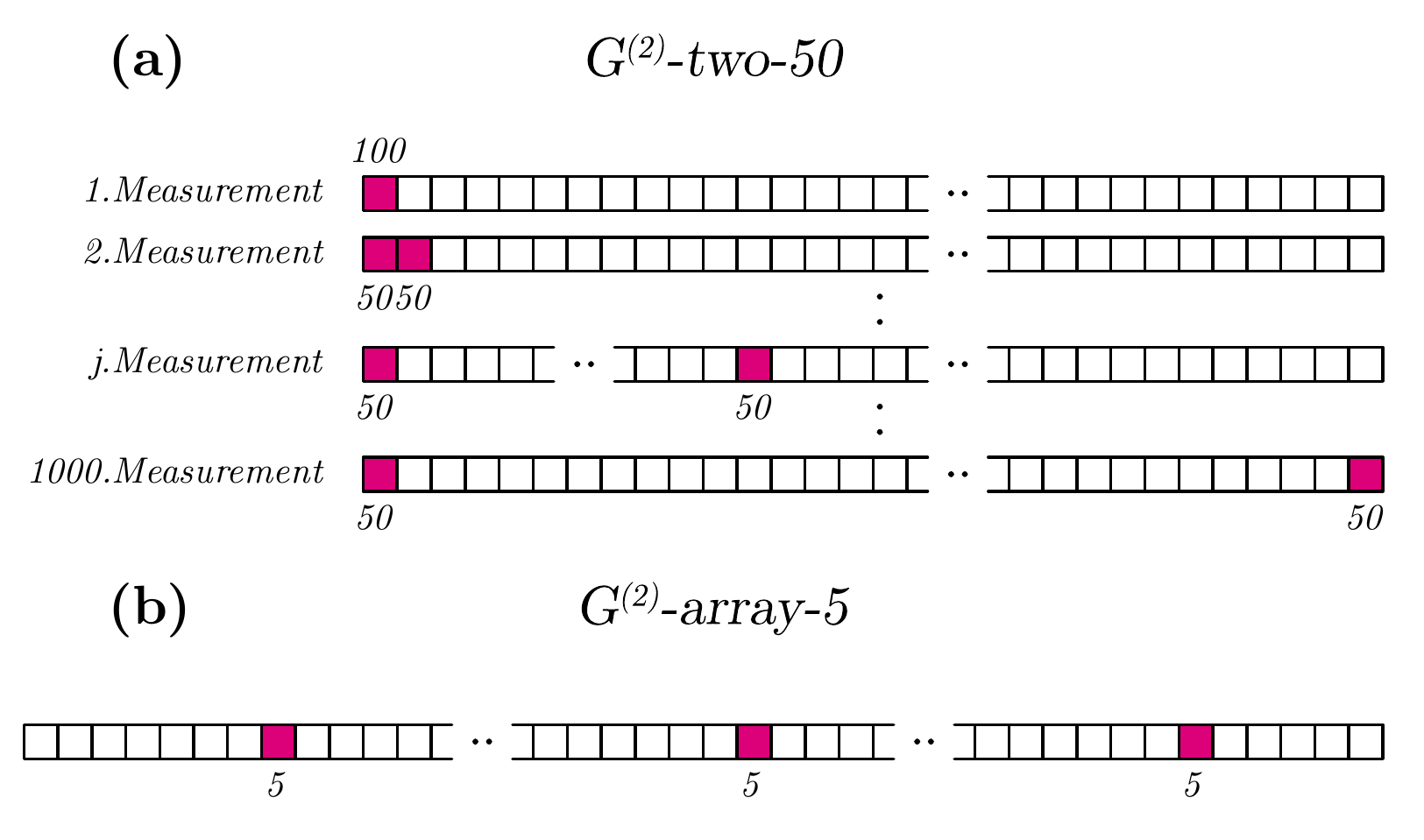}
 \caption{Sketch of the two different  $G^{(2)}$ detection schemes described in the main text. The squares indicate the different possible positions, colored squares correspond to positions at which detectors are located. The number next to a colored square denotes the number of detectors located at this particular position. (a) In the first \textit{$G^{(2)}$-two-50}-scheme, 50 detectors are fixed to position 1 of the 1000 possible positions spanning a length of 10\;km, whereas the second 50 detectors move from position to position. In the $j$th measurement the second 50 detectors are located at position $j$ ($j = 1, \ldots, 1000$). (b) In the second \textit{$G^{(2)}$-array-5}-scheme, the 100 detectors are split up into groups of 5 detectors at 20 different positions, distributed over the entire range of possible locations.}
  \label{fig:G2methods_scheme}
\end{figure}

Let $\mathcal{M}_{1}, \mathcal{M}_{2} \subset \lbrace 1,...,D\rbrace$ denote two subsets of detector positions. We define the $G^{(2)}$ observation vector specified by $\mathcal{M}_{1}, \mathcal{M}_{2}$ as
\begin{equation}
\bm{G}^{(2)}_{\mathcal{M}_{1}, \mathcal{M}_{2}}=\left(\left. I_{i,j}^{(2)}\right|_{i\in \mathcal{M}_{1}, j\in \mathcal{M}_{2}, i< j}\right)^\top\,,
\label{eq:obvector}
\end{equation}
where $I_{i,j}^{(2)}$ denotes the intensity correlation of detectors $i$ and $j$. For the explicit calculation of the lower bounds for the general  observation vector~\eqref{eq:obvector} we refer the reader to Appendix~\ref{sec:AppInt}.
Here, we assume $D=1\,000$ available detector positions spanning a total length of $10\,\mathrm{km}$. This is the maximum number of distinct positions given the size $A = 10\,\text{m}\times 10\,\text{m}$ of an individual telescope.\\
\indent In the following, we describe two different measurement schemes each utilizing the 100 detectors. In the first scheme, called \textit{$G^{(2)}$-two-50}-scheme, we suppose that we fix 50 detectors to the position $i=1$ ($\mathcal{M}_{1} = \lbrace 1 \rbrace$), and the remaining 50 detectors move from measurement to measurement to the positions $j=1$ to $j=1\,000$ [see Fig.~\ref{fig:G2methods_scheme} (a)]. In this case, for the $j$th measurement, the $G^{(2)}$ observation vector consists of only one element reading $G^{(2)}_{\mathcal{M}_{1},\mathcal{M}_2} = I^{(2)}_{1,j}$. We further assume that the measurements are independent of each other so that we can add the Fisher information as well as the lower bounds of the individual measurements. In the end, we divide the lower bound by the number of measurements to get an average precision per measurement. 

\begin{figure}[t!]
 \centering
 \includegraphics[width=.5\textwidth]{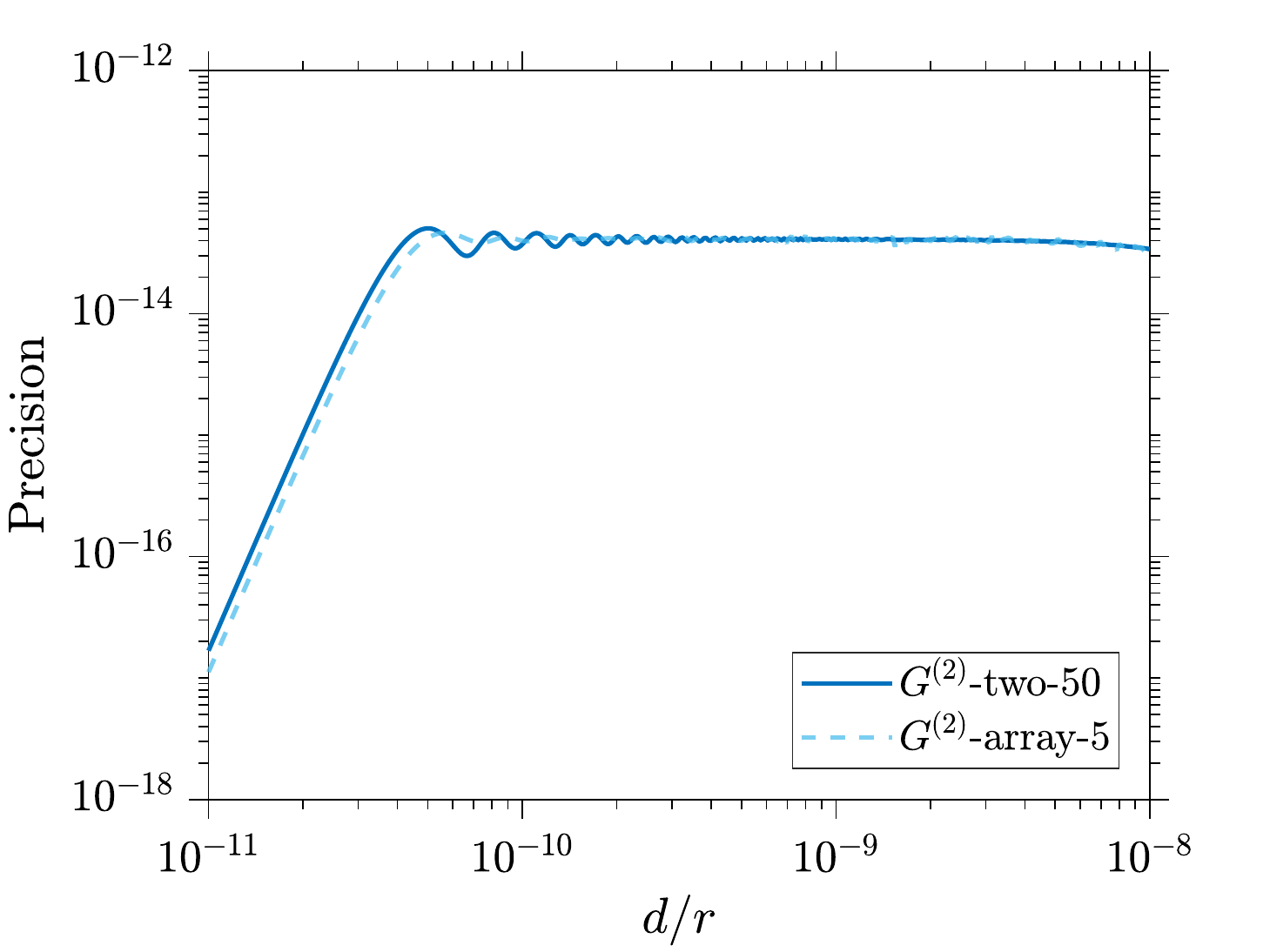}
 \caption{Precision of the source separation utilizing the two different  $G^{(2)}$ detection schemes,  calculated via the lower bound given in Appendix \ref{sec:AppInt}. We find that both methods display approximately the same amount of information about the separation of the two sources, leading to approximately the same precision. One can see that the $G^{(2)}$-array-5 curve is more smooth than the $G^{(2)}$-two-50 curve, since the measurement scheme for the former delivers multiple two-point correlations at once, whereas the measurement scheme for the latter only gives a single intensity correlation for each measurement.}
  \label{fig:G2methods}
\end{figure}

The second scheme, which we call \textit{$G^{(2)}$-array-5}-scheme, involves multiple positions within the array. Here, we assume that the 100 detectors are placed at 20 different positions, i.e., 5 detectors at each position in the array [see Fig.~\ref{fig:G2methods_scheme} (b)]. We denote the corresponding locations as $L_1,L_2,...,L_{20} \in \lbrace 1,...,D\rbrace$. In Fig.~\ref{fig:G2methods_scheme} (b), for instance, these locations are $L_1=8,...,L_{20}=995$. Considering all possible two-point correlations of the $20$ positions $L_1,L_2,...,L_{20}$ contained in the array, i.e., $\mathcal{M}_{1}=\mathcal{M}_{2}=\lbrace L_1,L_2,...,L_{20}\rbrace$, the observation vector reads
\begin{equation}
\bm{G}^{(2)}_{\text{Array}}=\left(I_{L_1,L_1}^{(2)},I_{L_1,L_2}^{(2)},...,I_{L_{20},L_{20}}^{(2)}\right)^\top\,.
\end{equation}
In this case, we divide the 1000 detector positions in intervals of $1000/20=50$ positions, whereby in each interval one detector position is populated with 5 detectors. We average the lower bound over 5 realizations, where for each realization the populated detector position of each interval is randomly chosen. \\
\indent In Fig.~\ref{fig:G2methods} we show for both $G^{(2)}$ detection schemes the calculated lower bound of the Fisher information matrix entry corresponding to the separation, directly related to the precision in estimating the separation of the two sources. We find that both $G^{(2)}$ measurement schemes achieve approximately the same precision. However, since the $G^{(2)}$-array-5 scheme provides multiple two-point correlations at once,  the curve is more smooth compared to the $G^{(2)}$-two-50 scheme, which only provides a single two-point correlation for each shot giving rise to small oscillations. 

\begin{figure*}
 \centering
 \includegraphics[width=\textwidth]{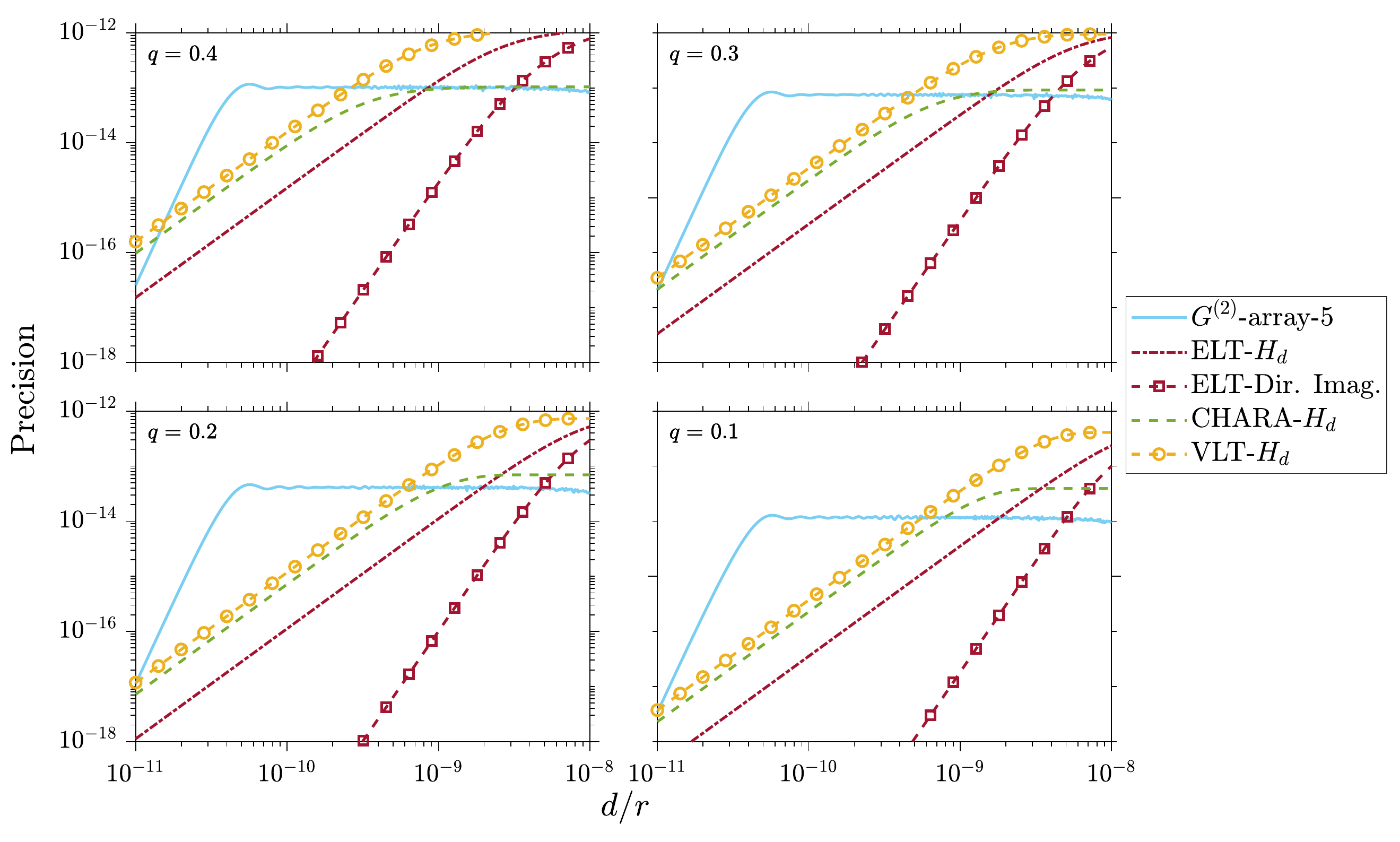}
 \caption{Calculated precision for the five different stellar imaging methods investigated in this paper against the angular separation $d/r$ of a double star for four different relative intensities  $q=0.1,0.2,0.3,0.4$. The solid (cyan) curve corresponds to the precision achievable via  $G^{(2)}$ intensity interferometry  ($G^{(2)}$-array-5 scheme) discussed in Sec.~\ref{sec:G2};
 the dashed-dotted (red) curve denotes the optimal $G^{(1)}$ method using a telescope of the size of ELT in combination with SPADE [Eq.~\eqref{eq:prec_gauss}]; the dashed-squared (red) curve corresponds to direct (Galilein) intensity imaging (Dir. Imag.) employing a telescope of the size of ELT  (see App. \ref{sec:AppDirect}); finally, the dashed (green) and the dashed-circle (yellow) curve correspond to the precision obtained via $G^{(1)}$ baseline amplitude interferometry [Eq.~\eqref{eq:prec_gauss}] as implemented at CHARA and VLT, respectively. The plots show that $G^{(2)}$ intensity interferometry is competitive to the different $G^{(1)}$ amplitude interferometric methods and even scales favourably for very small angular separations $d/r$ (for further details see main text).}
  \label{fig:StellarFIs}
\end{figure*}

\section{Comparison of $G^{(1)}$ and $G^{(2)}$ methods}\label{sec:comparison}
\noindent
In this section, we compare $G^{(2)}$ intensity interferometry as discussed  in Sec.~\ref{sec:G2}  with the various methods for  $G^{(1)}$ amplitude interferometry considered in Sec.~\ref{sec:G1}. For the comparison we use the four different relative source strengths $q=0.1,0.2,0.3,0.4$. 
In Fig.~\ref{fig:StellarFIs} we display the results obtained for the $G^{(2)}$-array-5 scheme, i.e., using an array of 100 detectors distributed at 20 different positions with a possible maximum baseline of 10\;km, and the outcomes of the four different interferometric methods discussed in Sec.~\ref{sec:G1}. The latter are: traditional $G^{(1)}$ amplitude interferometry using a single telescope with mirror size 39.3\;m as realized at the ELT; optimal $G^{(1)}$ amplitude interferometry using the same ELT-sized telescope in combination with SPADE; $G^{(1)}$ baseline amplitude interferometry using an array of telescopes the size 8.2\;m with a maximum baseline of 130.2\;m as realized at the VLT; $G^{(1)}$ baseline amplitude interferometry using an array of telescopes the size 1\;m with a maximum baseline of 330\;m as in CHARA. For all interferometric methods we assume a photon flux from the double star system of $2\cdot 10^{-2}$ photons per coherence time within an area of $10\,\mathrm{km}\times 10\,\mathrm{km}$ at a wavelength of $600\,\mathrm{nm}$; for equally bright sources ($q=\frac12$), this would mean that each source emits on average $10^{-2}$ photons per mode. The  telescope sizes, corresponding average  photon  numbers per telescope, baselines,  corresponding numerical  apertures,  and numbers of  telescopes  considered for each interferometer/telescope in Fig.~\ref{fig:StellarFIs} are listed in Table~\ref{tab:Comparison}.

Note that for the numerical simulations in Fig.~\ref{fig:StellarFIs} we assume an unrealistically small distance $r$ to the two sources (and as such unrealistic large numerical apertures) to avoid complications with very small numbers in the calculations; however, this does not change the relative position of the curves in Fig.~\ref{fig:StellarFIs}. 
%
%
%
In the figure, the precision for direct imaging using the ELT telescope is calculated via the procedure explained in App. \ref{sec:AppDirect} for the corresponding PSF, 
and the precision of the $G^{(2)}$ method is calculated via the lower bound given in App. \ref{sec:AppInt}.
\color{black}

\begin{table}[t!]
    \centering
    \begin{tabular}{c|c|c|c|c|c}
    method & \diameter$_{\text{tel}}$ & $N_{s,\text{tel}}$  & baseline & $\text{NA}$  & \#tel \\
    \hline
    $G^{(2)}$ & $10\,\mathrm{m}$ & $2\cdot 10^{-8}$ & $10\,\mathrm{km}$  & $10^{-8}$ & 100 \\
    ELT & $39.3\,\mathrm{m}$ & $3.1\cdot 10^{-7}$ & $-$ & $3.9\cdot 10^{-11}$ & 1 \\
    CHARA & $1\,\mathrm{m}$ & $2\cdot 10^{-10}$ & $330\,\mathrm{m}$ & $3.3\cdot 10^{-10}$ & 2 \\
    VLT & $8.2\,\mathrm{m}$ & $1.3\cdot 10^{-8}$ & $130.2\,\mathrm{m}$  & $1.3\cdot 10^{-10}$ & 2
    \end{tabular}
    \caption{Comparison of the different telescope sizes, average  photon  numbers per telescope, baselines,  corresponding numerical  apertures,  and numbers of  telescopes used for the numerical calculations displayed in Fig.~\ref{fig:StellarFIs}.}
    \label{tab:Comparison}
\end{table}


From Fig.~\ref{fig:StellarFIs} we can see that for the parameters of Table \ref{tab:Comparison}, $G^{(2)}$ intensity interferometry  is indeed competitive to $G^{(1)}$ amplitude interferometry. This is surprising, since $G^{(2)}$ measurements require two photons to be measured coincidentally, whereas $G^{(1)}$ measurements only need single photon detection; the precision  of $G^{(2)}$ intensity interferometry depends thus quadratically on the average photon number per mode, i.e., the degeneracy parameter $\delta$, whereas for  $G^{(1)}$ amplitude  interferometry the precision scales linearly with $\delta$. However, in both cases the precision also scales quadratically with the numerical aperture [see also the discussion in Sec.~\ref{sec:QFIsec}  following Eq.~\eqref{eq:prec_gauss}]. As in $G^{(2)}$ intensity interferometry the numerical aperture can exceed the numerical aperture of $G^{(1)}$ methods by a factor of $100$ or larger, this  can indeed compensate for the lower number of two-photon coincident detection events. 
In particular, as shown in Fig.~\ref{fig:StellarFIs}, when considering very small stellar separations (or feature sizes), we find a regime of angular separations $d/r$ for which the precision of the $G^{(2)}$ method clearly outperforms the one of the $G^{(1)}$ methods. Here, the advantage of the much larger numerical apertures achievable in $G^{(2)}$ intensity interferometry compared to $G^{(1)}$ amplitude interferometry  clearly overcomes the drawback of the quadratic dependency on the small degeneracy parameter  $\delta$ at optical frequencies. Yet, the slope of the $G^{(2)}$ precision drop-off is steeper than the one of the $G^{(1)}$ precision as the latter depends only linearly on $\delta$.

\section{Possible Implementations of VLBII}\label{sec:implementation}
\noindent
Currently, in astronomical  imaging,  both amplitude interferometry and intensity interferometry are employed. Direct (Galilean) imaging is presently pushed to the limit with the ELT  in  Chile, having a compound mirror of 39.3 m in diameter \cite{ELT_Construction_2014}. Baseline amplitude interferometry is  realized  by GRAVITY at VLT and CHARA using  baselines of   \SI{130.2}{\m} \cite{2018_VLT} and \SI{330}{\m} \cite{2020_Anugu}, respectively. Recently, intensity interferometry was implemented with 4 telescopes by the  VERITAS collaboration at a maximum baseline of \SI{172.5}{\m} \cite{2020_Kieda}; already beforehand, spatial  intensity correlations have been measured with starlight using smaller baselines \cite{2018_GuerinKaiser_SpatialStellarII,2020_Acciarri_spatialI_MAGIC}. While so far the baselines utilized in intensity interferometry have been of reasonable size, interesting prospects arise at baselines of the order of \SI{2}{\km} or even \SI{10}{\km}. Both will allow for sub-milli-arcsecond resolution. Presently, the Cherenkov Telescope Array (CTA) is studied for this purpose as a possible candidate for VLBII aiming for a baseline of  \SI{2.4}{\km} \cite{CTASouthObservatory2021,Dravins2013,2019_Kieda}

Intensity interferometry on stars as faint as apparent magnitude 5 are estimated to be possible with  $G^{(2)}$ setups using currently existing Cherenkov telescopes with diameters on the order of \SI{4}{\m} to \SI{23}{\m} \cite{2020_Kieda}. The future CTA is projected to allow intensity interferometric observations of stars even as faint as apparent magnitude 7 \cite{2012_Nunez}. At these apparent magnitudes around 2700 (magnitude 5) and 25000 (magnitude 7) astronomical targets can be selected for observation \cite{SIMBAD}, however not all of them will be observable from a given location on earth or emitting into the required optical bands. Only about 2600 objects from the bright star catalogue can be selected for VLBII observation when additionally including spectral limitations \cite{Dravins2013, Hoffleit1995}. Also implementing third-order intensity interferometry at CTA seems possible, and would aid the image reconstruction process \cite{Malvimat2013}. 
 
At baselines of the order of \SI{10}{\kilo \m}, angular resolutions on the order of $\sim 5 \,\upmu\text{as}$ are possible for $G^{(2)}$ setups at visible wavelengths. While such high angular resolutions are probably not required for diameter measurements of main sequence stars brighter than  apparent magnitude 8 (having typical angular resolutions not far below $\sim 0.1 \, \text{mas}$ \cite{2017_Stee}) they can be utilized for a number of other interesting astronomical measurements. Resolutions of $\sim 0.05 \, \text{mas}$ are required to image dark or bright spots of stars within the milky way \cite{2012_Nunez}, measure stellar deformities due to rotation \cite{2015_Nunez}, circumstellar disks and hot star winds \cite{Dravins16}, or properties of interacting binary stars \cite{2012_Dravins}. Even higher angular resolutions are beneficial if stellar images should be reconstructed for small and faint stars. These ultra-high angular resolutions are also needed if exoplanet transits at distant stars are to be studied, with  exoplanets typically having an angular extent 20 times smaller than the star \cite{2017_Stee}. Additionally, these high resolutions might facilitate the study of otherwise unresolved features in distant galaxies or distant supernovae.
 
Observations of pre-series stars in their formation and evolved stars up to planetary nebulae has been proposed for amplitude interferometry \cite{2017_Stee}. According to our estimations, this should be achievable with VLBII, yet utilizing presumably much simpler setups. This implies that nearly all stages in the life of a star would be observable with VLBII at higher resolutions than with amplitude interferometry considering realistic setups. 

\section{Conclusions}\label{sec:discussion}
\noindent
In conclusion, we have presented a quantitative comparison of the precision achievable in stellar astronomy, using $G^{(1)}$ amplitude interferometry on the one hand and  $G^{(2)}$ intensity interferometry on the other hand for the benchmark problem of determining the separation between two adjacent stars. The comparison is based on well-established  measures  from  estimation  theory, the Fisher and the quantum Fisher information. The two quantities allow for the determination of the  amount  of  information, i.e., the precision, that one can achieve in a given measurement (Fisher information) or even in an optimal measurement (quantum  Fisher information) of the electromagnetic field for estimating the centroid $z_0$, the separation $d$ and the relative strength $q$ of a double star system. After a short introduction to the Fisher and the quantum Fisher information in Sec.~\ref{sec:QFI}, we investigated in  Sec.~\ref{sec:G1} various imaging methods which make use of $G^{(1)}$ amplitude interferometry. The latter included  direct (Galilean) imaging  as realized at the  ELT in Chile, optimal  imaging as could be implemented at the ELT by use of SPADE,  and another optimal imaging method using baseline $G^{(1)}$ interferometry as implemented, e.g., in the GRAVITY collaboration and at CHARA. In Sec.~\ref{sec:G2}, we determined the precision which one can achieve via $G^{(2)}$ measurements by use of Very Large Baseline Intensity  Interferometry (VLBII), employing real parameters as implemented, e.g., at the southern  Cherenkov Telescope Array (CTA) in Chile. 

In Sec.~\ref{sec:comparison}, we compared the two methods and found
that for realistic parameters VLBII can indeed be competitive
and even outperform $G^{(1)}$ amplitude interferometry. The reason is that even though $G^{(2)}$ intensity interferometry
scales quadratically with the small degeneracy
parameter $\delta$ while $G^{(1)}$ amplitude interferometry
scales linearly with $\delta$, both methods depend quadratically
on the numerical aperture. Since for VLBII the
latter can be orders of magnitude larger compared to realistic
$G^{(1)}$ systems, the precision in $G^{(2)}$-measurements
can be of comparable value and even overcome the one
of $G^{(1)}$ amplitude interferometry, in particular when considering
very small stellar separations or feature sizes of
the object of interest.

More precisely, calculating the quantum Fisher information matrix
and analyzing the corresponding analytic expression
for the precision about the separation of the two
stars, we learned that for optimal $G^{(1)}$ measurements the precision
drops off to zero in the limit of infinitesimally
small separations, depending on the numerical aperture. I.e., for larger
numerical apertures the drop-off starts at smaller angular
separations $d/r$ [see the discussion following Eq. \eqref{eq:prec_gauss}].
This statement
holds true in a similar fashion for $G^{(2)}$ measurements (even though the detection
involves two photons what steepens the slope of the
drop-off). Fig.~\ref{fig:StellarFIs} shows that the different onsets of these reductions  lead to a regime of very
small separations of the double star system for which the
precision achievable with $G^{(2)}$  intensity interferometry is
higher than that of $G^{(1)}$  amplitude interferometry.


In the future, both optimal $G^{(1)}$ amplitude interferometry and VLBII can be pushed to increase performance. For example, it is possible to use quantum entangled networks to further increase the numerical aperture in $G^{(1)}$ amplitude interferometry  by teleporting collected photons over large distances \cite{Gottesman12}. Teleportation preserves the relative phase of the photons, and photon transmission losses can be side-stepped by prior established photonic entanglement.
VLBII with decent baselines was recently established, but remain at the testing stage \cite{2018_GuerinKaiser_SpatialStellarII,2020_Kieda,2020_Acciarri_spatialI_MAGIC}. Larger arrays like CTA  have been considered an ideal implementation of VLBII, with  prospective  baselines of about  $\sim \SI{2}{\kilo \m}$ and joint recovery of many spatial frequencies, utilizing about $100$  telescopes at once \cite{2019_Kieda}, which would enable the reconstruction of conventional images \cite{2012_Nunez,2015_Nunez}. VLBII results might be further improved by exploring optimal detector array geometries or exploiting higher-order spatial correlations \cite{Oppel2012}. Envisioning VLBII with $100$ telescopes of diameter $\sim \SI{10}{\m}$ and baselines on the order of $\sim \SI{10}{\kilo \m}$, resolutions of $\sim 0.005 \, \text{mas}$ are possible, surpassing the resolution of current and planned amplitude interferometers by a factor of 20 or more, and even surpassing the  resolution of  the EHT Collaboration.

\section*{Acknowledgements}
\noindent M.B., S.R., and J.v.Z.\ gratefully acknowledge funding and support by the International Max Planck Research School - Physics of Light. Z.H.\ and P.K.\ are supported by the EPSRC Quantum Communications Hub, Grant No.\ EP/M013472/1, and the EPSRC grant Large Baseline Quantum-Enhanced Imaging Networks, Grant No.\ EP/V021303/1.

\begin{appendices}

\section{Direct imaging precision}
\label{sec:AppDirect}
\noindent Similar to Eq.~\eqref{eq:densitytm}, we write the one-photon truncated state (neglecting the vacuum) as
\begin{align}
    \rho = q \ket{\Psi_+}\bra{\Psi_+} + (1-q) \ket{\Psi_-}\bra{\Psi_-}\,,
\end{align}
where 
\begin{align}
    \ket{\Psi_\pm} = \int dz\, \Psi_\pm(z) \ket{z}\,.
\end{align}
Here, $q$ characterizes the relative source strength and $z$ is the image plane coordinate. Assuming unit magnification $\Psi_-(z)=\braket{z-z_1|\Psi}$ denotes the state of a photon originating from the source at $z_1$, $\Psi_+(z)=\braket{z-z_2|\Psi}$ denotes the state of a photon originating from the source at $z_2$. The probability to detect a photon at position $z$ is then given by
\begin{align}
\label{eq:probdirect}
    p(z) = \braket{z|\rho|z} = q |\Psi_+(z)|^2 + (1-q) |\Psi_-(z)|^2\,.
\end{align}
The multi-parameter Fisher information matrix of Eq.~\eqref{eq:fi} for direct imaging is then calculated with the probability distribution of Eq.~\eqref{eq:probdirect} with respect to the three parameters $\theta_j$, $j\in\lbrace 1,2,3\rbrace$, given by the centroid $z_0$, the separation $d$ and the relative source strength $q$. By numerically calculating the inverse of the Fisher information matrix and taking the inverse of the diagonal element corresponding to the separation we find the precision bound for direct imaging in terms of Eq.~\eqref{eq:clVarCRB}.

\section{Intensity interferometry precision}\label{sec:AppInt}
\noindent Consider a general vector random variable $\bm{X}=(X_1,...,X_M)^\top\in\mathbb{R}^M$ whose probability density $p_{\bm{X}|\bm{\Theta}}(\bm{x}|\bm{\theta})$ depends on $\bm{\theta}$. The Fisher information $\mathcal{F}(\bm{\theta})$ of $\bm{X}$ on $\bm{\theta}$ is bounded from below by \cite{stein2014lower}
\begin{equation}
\mathcal{F}(\bm{\theta})   \succeq \left(\frac{\partial \boldsymbol{\mu}}{\partial \bm{\theta}}\right)^\top\bm{C}^{-1}\left(\frac{\partial \boldsymbol{\mu}}{\partial \bm{\theta}}\right)\,,
\label{eq:TLS1.14}
\end{equation}
where $\bm{\mu}=(\braket{ X_1}(\boldsymbol{\theta}),...,\braket{ X_M}(\boldsymbol{\theta}))^\top$ is the mean observation vector of the random vector $\bm{X}$ and ${\bm{C}=\braket{(\bm{X}-\bm{\mu})(\bm{X}-\bm{\mu})^\top}(\boldsymbol{\theta})}$ is the covariance matrix of $\bm{X}$. In what follows, we specify the random vectors for second-order intensity correlation measurements and calculate the corresponding expectation values.\\
\indent The correlation function $G^{(2)}(\mathbf{r}_1,\mathbf{r}_2)$ at positions $\mathbf{r}_1$ and $\mathbf{r}_2$ is defined as
\begin{equation}
\begin{split}
    G^{(2)}(\mathbf{r}_1,\mathbf{r}_2)&=\braket{ E^\dagger(\mathbf{r}_1)E^\dagger(\mathbf{r}_2)E(\mathbf{r}_2)E(\mathbf{r}_1)}\\
    &=\mathrm{Tr}[E(\mathbf{r}_2)E(\mathbf{r}_1)\rho E^\dagger(\mathbf{r}_1)E^\dagger(\mathbf{r}_2)]\\
    &=\mathbb{E}_{\bm{A}}\left[\vert\psi_{\bm{A},d}(\mathbf{r}_1)\vert^2 \vert\psi_{\bm{A},d}(\mathbf{r}_2)\vert^2\right]\,,
\label{eq:G2A}
\end{split}
\end{equation}
where the optical equivalence theorem was used in the last line of Eq.~(\ref{eq:G2A}). Here, $\mathbb{E}_{\bm{A}}[.]$ denotes the expectation value with respect to the source amplitudes. Since the sphere is partitioned into specific detector positions, the positions $\mathbf{r}_1$ and $\mathbf{r}_2$ are replaced by two detector indices $i$ and $j$ with detector areas
\begin{equation}
\mathcal{A}_{i/j}=\lbrace (\varphi,\theta):-\tilde{\varphi}\leq \varphi< \tilde{\varphi},l_{i/j}\leq \theta\leq r_{i/j}\rbrace\,,
\label{eq:pixelarea}
\end{equation}
where $\vert r_{i/j}-l_{i/j}\vert=2\tilde{\theta}/D$ and $\tilde{\theta}$ denotes the angle of the numerical aperture. Then the correlation function of a single shot becomes an intensity correlation $I_{i,j}^{(2)}$ of detectors $i$ and $j$. Thus the random vectors for the second-order intensity correlation measurements are the ones specified in Eq.~\eqref{eq:obvector}. We now evaluate the corresponding expectation values.\\
\indent We assume that the photons are freely propagating, i.e., their wave function can be described as a spherical wave
\begin{equation}
    \psi(\mathbf{r})=\frac{1}{\sqrt{4\tilde{\varphi}\sin\tilde{\theta}}}\frac{e^{ikr}}{r}\,.
\end{equation}
We then have for a displaced emission at point $\bm{z}$ the wave function
\begin{equation}
    \psi(\mathbf{r}-\bm{z})=\frac{1}{\sqrt{4\tilde{\varphi}\sin\tilde{\theta}}}\frac{e^{i k|\mathbf{r}-\bm{z}|}}{|\mathbf{r}-\bm{z}|}\,.
\end{equation}
In the far field we can approximate $|\mathbf{r}-\bm{z}|\approx |\mathbf{r}|=r$ in the denominator. In the phase we need a more sensitive approximation as $|\mathbf{r}-\bm{z}|\approx r-\mathbf{r}\bm{z}/r=r-z\cos\theta$. In what follows we neglect the $1/r$ dependence and the constant phase coming from the first term in $r-z\cos\theta$. Therefore, the states can be written in far-field approximation as
\begin{equation}
    \psi(\mathbf{r}-\bm{z}_{1/2})=\frac{1}{\sqrt{4\tilde{\varphi}\sin\tilde{\theta}}}e^{-ik(z_0\mp d/2)\cos\theta}\,.
\end{equation}
\indent Now, considering the finite detector size Eq.~\eqref{eq:pixelarea} the mean values of the conditional intensity correlations $I_{i,j|\bm{A}}^{(2)}$ yield
\begin{equation}
\begin{split}
    \mu_{i,j|\bm{A}}&=\int_{\mathcal{A}_{i}} d^2r_1 \int_{\mathcal{A}_{j}} d^2r_2\, \vert\psi_{\bm{A},d}(\mathbf{r}_1)\vert^2 \vert\psi_{\bm{A},d}(\mathbf{r}_2)\vert^2\\
    &=\eta_{i|\bm{A}}\eta_{j|\bm{A}}
\end{split}
\end{equation}
with 
\begin{equation}
\begin{split}
    \eta_{p|\bm{A}}\coloneqq&\,\int_{\mathcal{A}_{p}} d^2r\,\vert\psi_{\bm{A},d}(\mathbf{r})\vert^2\\
    =&\,\left| A_+\right|^2\alpha_p+\left| A_-\right|^2 \alpha_p+A_+^*A_-\beta_p+A_+A_-^*\beta_p^*\,,
\end{split}
\end{equation}
where 
\begin{align}
\begin{split}
    \alpha_p\coloneqq&\,\int_{\mathcal{A}_p} d^2r\,\vert\psi(\mathbf{r}-\bm{z}_2)\vert^2\\
    =&\,\,\frac{1}{2\sin\tilde{\theta}}\left[\cos(l_p)-\cos(r_p)\right]\,,
\end{split}\\
\begin{split}
    \beta_p\coloneqq&\,\int_{\mathcal{A}_p} d^2r\,\psi^*(\mathbf{r}-\bm{z}_2)\psi(\mathbf{r}-\bm{z}_1)\\
    =&\,\,\frac{1}{2\sin\tilde{\theta}ikd}\left[e^{ikd\cos(l_p)}-e^{ikd\cos(r_p)}\right]\,.
\end{split}
\end{align}
Using Eq.~\eqref{eq:G2A} and the following identities, fulfilled by the source amplitudes $A_+$ and $A_-$
\begin{gather}
\mathbb{E}_{\bm{A}}[A_+^*A_-]=\mathbb{E}_{\bm{A}}[A_-^*A_+]=0\,,\\
\mathbb{E}_{\bm{A}}[A_+ A_-]=\mathbb{E}_{\bm{A}}[A_- A_+]=0\,,\\
\mathbb{E}_{\bm{A}}[\left| A_+\right|^2]= N_s q\,,\mathbb{E}_{\bm{A}}[\left| A_-\right|^2]=N_s(1-q)
\end{gather}
the expectation values $\mu_{i,j}=\mathbb{E}_{\bm{A}}[\mu_{i,j|\bm{A}}]$ can be calculated, which can then be used to find the derivatives with respect to the unknown parameters. Analogously the elements of the covariance matrix can be calculated via
\begin{equation}
C_{i,j,k,l}=\mathbb{E}[I_{i,j}^{(2)}I_{k,l}^{(2)}]-\mu_{i,j}\mu_{k,l}\,.
\end{equation}
We are then able to compute numerically the lower bound Eq.~\eqref{eq:TLS1.14} for the different measurement schemes characterized in the main text.
\end{appendices}

\bibliography{g2}

\end{document}